\documentclass[10pt, final, journal, a4paper, oneside]{IEEEtran}
\usepackage{amsmath,amsfonts}
\usepackage{algorithmic}
\usepackage{algorithm}
\usepackage{array}
\usepackage[caption=false,font=normalsize,labelfont=sf,textfont=sf]{subfig}
\usepackage{textcomp}
\usepackage{stfloats}
\usepackage{url}
\usepackage{verbatim}
\usepackage{graphicx}
\usepackage{cite}
\usepackage{tikz}
\usepackage[T1]{fontenc}
\usepackage{bbm}
\usepackage{dsfont}
\usepackage{amssymb}
\usepackage{orcidlink}

\usepackage{array}
\usepackage{multirow}

\usepackage{hyperref}
\hypersetup{
    colorlinks = true,
    linkcolor = orange,
    citecolor = purple,
    filecolor = black,      
    urlcolor = teal,
}

\begin{document}
\title{Influence of noise in entanglement-based quantum networks}

\author{Maria Flors Mor-Ruiz$^{\orcidlink{0000-0003-4921-5929}}$ and Wolfgang D\"ur$^{\orcidlink{0000-0002-0234-7425}}$ 
\thanks{Manuscript received 5 May 2023; revised 29 November 2023; accepted 20 December 2023. Date of publication 25 March 2024; date of current version 19 June 2024. This research was funded in whole or in part by the Austrian Science Fund [Fonds zur F\"orderung der wissenschaftlichen Forschung (FWF)] 10.55776/P36010 and 10.55776/P36009 and by the Europ\"aischen Union - NextGenerationEU. \textit{(Corresponding authors Maria Flors Mor Ruiz; Wolfgang D\"ur.)}}
\thanks{The authors are with the Universit\"at Innsbruck, Institut f\"ur Theoretische Physik,  Technikerstra{\ss}e 21a, 6020 Innsbruck, Austria (email: maria.mor-ruiz@uibk.ac.at).}
\thanks{This paper has been published in IEEE Journal on Selected Areas in Communications, vol. 42, no. 7, pp. 1793-1807, July 2024, DOI: 10.1109/JSAC.2024.3380089.}
\thanks{\copyright IEEE 2023. Personal use of this material is permitted. Permission from IEEE must be obtained for all other uses, in any current or future media, including reprinting/republishing this material for advertising or promotional purposes, creating new collective works, for resale or redistribution to servers or lists, or reuse of any copyrighted component of this work in other works.}}

\maketitle
\thispagestyle{empty}

\begin{abstract}
We consider entanglement-based quantum networks, where multipartite entangled resource states are distributed and stored among the nodes and locally manipulated upon request to establish the desired target configuration. Separating the generation process from the requests enables a pre-preparation of resources, hence a reduced network latency. It also allows for an optimization of the entanglement topology, which is independent of the underlying network geometry. We concentrate on establishing Bell pairs or tripartite GHZ states between arbitrary parties. We study the influence of noise in this process, where we consider imperfections in state preparation, memories, and measurements - all of which can be modeled by local depolarizing noise. We compare different resource states corresponding to linear chains, trees, or multi-dimensional rectangular clusters, as well as centralized topologies using bipartite or tripartite entangled states. We compute the fidelity of the target states using a recently established efficient method, the noisy stabilizer formalism, and identify the best resource states within these classes. This allows us to treat networks of large size containing millions of nodes. We find that in large networks, high-dimensional cluster states are favorable and lead to a significantly higher target state fidelity.
\end{abstract}


\section{Introduction} \label{sec:introduction}
\noindent
\IEEEPARstart{Q}{uantum} networks \cite{Kimble2008,wehner_internet, Azuma2021, azuma_2022} form one of the pillars of upcoming quantum technologies and allow one to push promising applications to distributed settings. While quantum computers and quantum metrology for themselves already provide exciting and relevant applications \cite{Acin2018Quantum, Riedel2019Europe} in multiple branches of science and technology, connecting quantum devices in a network adds additional possibilities and features \cite{Cacciapuoti2020, Eldredge2018, Sekatski2020, CiracDistributed, caleffi_2018}. In addition, one of the key applications of quantum networks is in the context of cryptography, to establish secret keys in bipartite communications settings \cite{Gisin2002, ShorSimple}, for conference key agreement \cite{Murta2020Quantum, hahn_anonymous} or for applications in secret sharing \cite{markham_graph, Hillery1999} and secret voting \cite{Hillery2006, Vaccaro2007}. These applications are based on entangled quantum states shared among some network nodes. Establishing such entangled states over long distances has been identified as one of the crucial challenges in this context. This also solves the quantum communication issue, as entanglement can be used by means of teleportation \cite{BennetTeleporting} to transmit arbitrary quantum information between nodes and form distributed sensor networks or connect distributed quantum computers. Quantum repeaters \cite{Briegel1998, Sangouard2011, Munro2015} allow for an efficient distribution of entangled states, thereby overcoming the exponential scaling of resources due to loss and imperfections. Promising experimental realizations of entanglement distribution, quantum repeaters, and quantum teleportation have already been performed in different physical platforms \cite{Krutyanskiy2023, Pompili2021, chen2021integrated, li2019experimental, hermans2022qubit, langenfeld2021quantum}, where entanglement was generated over distances up to 50 km on the surface of the earth and over 4,600 km in space-to-ground, and memory qubits with intrinsic dephasing time up to seconds were demonstrated. 

\begin{figure}[!t]
    \centering
    \includegraphics[width=\columnwidth]{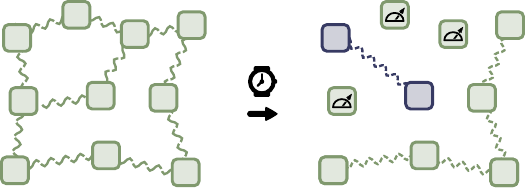}
    \caption{Schematic depiction of an entanglement-based quantum network, where squares represent nodes and snake-like lines represent entanglement between them. On the left, the distributed resource state is depicted with fragmented entanglement lines to indicate that the state is noisy. The state is stored until a request enters the network, for instance, a bipartite connection between two nodes. This is fulfilled by manipulating the resource state via local operations, e.g., measurements, as on the right side. Both storage and manipulation introduce further noise, thus the entanglement lines on the right are more fragmented.}
    \label{fig:introduction}
\end{figure}

Much of the experimental effort is still focused on establishing the building blocks for quantum repeaters, including long-time quantum memories or quantum interfaces and demonstrating point-to-point communication. However, in recent years theoretical effort has increased to set the stage for quantum networks consisting of multiple nodes, going beyond point-to-point communication \cite{wehner_internet, CacciapuotiWhen, pirker_quantum_2019, vanmeter_recursive, Muralidharan2016Optimal, bugalho2023distributing, patil2022entanglement, fischer_2021, pant2019routing}. Most of the approaches \cite{matsuzaki_2010, vanmeter_recursive, epping2016large, wehner_internet, pirandola2019end, wallnofer_simulating_2022, kozlowski_2019, coutinho2022robustness} thereby follow a bottom-up strategy, resembling the ones successfully employed in classical networks. Entanglement is established among requests, and entanglement routing, together with protocols to maintain or improve its quality, are among the key issues. While strategies to enable long-distance quantum communication are in principle known and scale efficiently with the distance, still practical challenges remain and protocols involve a large overhead in resources in terms of used memory and in particular required time \cite{Azuma2015All, coopmans_netsquid_2021, wallnofer_simulating_2022, Pirandola2017Fundamental, Munro2015, Sangouard2011, Guha2015, Sanogouard2009}. The fact that many of the involved processes, such as photon transmission or entanglement purification to increase state fidelities, are probabilistic (or require huge, inaccessible spatial resources such as multiple parallel channels or large memories), makes the production of long-distance entanglement costly. This leads to relatively low rates and hence large latency in the network. In particular, if in a large network, the process of producing entanglement between some of the nodes only starts after a request is received, there is a significant waiting time until the process can be concluded, and entanglement can be used.

In \cite{pirker_modular_2018, pirker_quantum_2019, meignant_2019, gyongyosi2019opportunistic}, an alternative approach towards quantum networks was put forward, which makes use of the unique features of quantum entanglement and does not have a classical counterpart. Since entanglement is a resource, it can be established even before a request arrives and the entangled qubits are stored in quantum memories until needed. Taking into account all possible requests by storing all kinds of bipartite entangled states is, however, too costly and impractical. Nevertheless, what one can do, is to establish and store some multipartite entangled resource states, which are chosen in such a way that memory requirements are minimized while full functionality can be guaranteed. The functionality of a network is thereby defined as the set of desired target configurations of states, i.e., defining the set of parties that should share certain entangled states. One natural desired functionality can be, e.g., to share a maximally entangled Bell pair between any two parties, where the requests correspond to the specification of these two parties. The resource states are therefore manipulated locally, either by some unitary operations or by measurements, to transform them to the desired target configuration. In the given example, one may either store all possible combinations of $N(N-1)/2$ Bell pairs and use the required one, for an $N$-node network, but also a single copy of a so-called GHZ state suffices to fulfill the task \cite{miguel-ramiro_optimized_2021}. The latter only requires the storage of a single qubit per node, corresponding to a quadratic reduction in required memory size. In \cite{miguel-ramiro_optimized_2021} a method to identify optimized resource states with minimal storage requirement was put forward, which provides one with the entanglement topology for a given desired network functionality. Remarkably, this entanglement topology is independent and separated from the underlying network geometry, at least from the perspective of storage and adaption to the desired target state. Clearly, for generating the resource state, the geometry is relevant, however, this process is separated from the adaption of the resource state and can be done beforehand, i.e., before the request arrives and the network otherwise lies idle.

In this paper, we concentrate on the influence of noise and imperfections in such entanglement-based quantum networks (EBQNs). All involved processes - the generation of entangled resource states, their storage in a quantum memory, and the manipulation by means of local unitary operations and measurements, are noisy in practice, and all these imperfections crucially influence the quality of the generated target states, as schematically depicted in Fig.~\ref{fig:introduction}. For some specific states and configurations such as an $N$-qubit GHZ state \cite{dur_stability_2004, hein_entanglement_2005}, an entanglement switch\footnote{We refer to \textit{entanglement switch} as a network structure that performs entanglement swapping. In literature, this is also referred to as a quantum switch, which, however, also refers to an approach for communication in indefinite causal orders \cite{caleffi_2020, chiribella2021indefinite, caleffi2023beyond}.} \cite{Cuquet2012, coopmans_netsquid_2021, vardoyan_stochastic_2019, Vardoyan2020, nain_2020, nain2022analysis}, the six-qubit Butterfly state \cite{epping_robust_2016, leung_butterfly, hahn_quantum_2019, Satoh2016} or for one-dimensional resource states \cite{jong_2022} the entanglement properties have been studied to some extent, however, the influence of noise in general EBQNs is largely unexplored. While previous work concentrated on memory requirements \cite{pirker_modular_2018,miguel-ramiro_optimized_2021}, it is known that different kinds of states suffer differently under noise \cite{dur_stability_2004, hein_entanglement_2005, Frowis2018}. For instance, a GHZ state is particularly susceptible to noise, and even though it is a resource with minimal storage requirements, it will provide a poor performance in the presence of noise, imperfections, and decoherence for large systems. In particular, one cannot produce entangled Bell pairs from GHZ states if the noise is too large, and the threshold becomes more and more stringent with system size \cite{dur_stability_2004}. 

We investigate different classes of entangled states and analyze their performance in the presence of noise. We model the influence of imperfect state preparation, imperfect memory, and noisy measurements using local depolarizing noise acting on each of the qubits stored in different nodes. Given that each qubit sees an independent environment, this is an adequate and sufficiently general error model. We concentrate on network requests corresponding to the generation of a single Bell pair or a single three-qubit GHZ state, where the request corresponds to the specification of the target parties. We use so-called graph states \cite{hein_multiparty_2004, hein_entanglement_2006} as resource states, where the graph topology directly corresponds to its entanglement topology. We consider quantum networks of fixed size $N$ based on resource graph states corresponding to linear chains, trees, or multi-dimensional rectangular grids or clusters, as well as collections of bipartite or tripartite entangled states arranged in a centralized switch topology.

Our main findings can be summarized as follows:
\begin{itemize}
\item For all these configurations, we provide formulas to compute the fidelity of target states in the low noise regime and analytic formulas for several of these settings using the recently introduced noisy stabilizer formalism (NSF) \cite{mor_noisy}. This allows us to assess the suitability of different resource states.
\item For an entanglement switch configuration with one central node, we show that three-qubit GHZ states perform better than Bell pairs when the two target nodes have qubits that are part of the same three-qubit GHZ state. 
\item We compare the achievable fidelities and error thresholds and identify optimal states among the considered classes. We find that trees and high-dimensional cluster states are favorable for large-scale networks. For trees, however, the entanglement structure is largely destroyed, while this is not the case in high-dimensional clusters.  
\end{itemize}
We emphasize that we can treat large resource states containing millions of qubits in an exact way, which is made possible by using the stabilizer formalism to describe states and their manipulation, together with the NSF to treat the influence of noise. For general states, the effort would scale exponentially with the number of qubits. 

The paper is organized as follows. In Sec.~\ref{sec:background} we provide some background information on EBQNs, graph states, and their manipulations, as well as the NSF. We describe the problem setting in Sec.~\ref{sec:setting}, where we also discuss the noise model and the resource and target states we consider. We consider a single Bell pair as a target in Sec.~\ref{sec:fidelity} and Sec.~\ref{sec:BP}, where we first compute a general expression for the fidelity and later we analyze all resource states, which are compared in Sec.~\ref{sec:comparison}. A similar analysis for three-qubit GHZ states is performed in Sec.~\ref{sec:3GHZ}. We discuss other target states briefly, summarize, and conclude in Sec.~\ref{sec:conclusion}.  

\section{Background} \label{sec:background}
In this section, we give a brief overview of EBQNs, some basic notations, and results concerning graph states, followed by a brief description of the NSF, which is used throughout this paper. Table~\ref{tab:notation} compiles the functions and abbreviations used throughout the text.

\begin{table}[!t]
\caption{Summary of used functions and abbreviations in order of appearance.
\label{tab:notation}}
\centering
\begin{tabular}{|l|l|}
\hline
\textbf{Function/Abbreviation} & \textbf{Meaning} \\ \hline \hline
EBQN & Entanglement-Based Quantum Networks \\
NSF & Noisy Stabilizer Formalism \\
LER & Low Error Regime \\ 
$g(x)$ & $[1-(-1)^x]/2$ \\
1D & one-dimensional \\ 
PBC & Periodic Boundary Conditions \\
2D & two-dimensional \\ 
$k$D & $k$-dimensional, where $k\in\mathbb{N}^+$ \\ \hline
\end{tabular}
\end{table}

\subsection{Entanglement-based quantum networks}\label{ssec:EBQN}
In recent years, the study of quantum networks relying on quantum repeaters has received large attention \cite{matsuzaki_2010, vanmeter_recursive, epping2016large, wehner_internet, pirandola2019end, wallnofer_simulating_2022, kozlowski_2019}, where quantum repeaters refresh the entanglement to counteract the influence of noise and decoherence. These works have in common that resources in the network are generated on demand, and we refer to them as the bottom-up approach to quantum networks. To establish these resources, network devices must perform routing tasks \cite{epping_robust_2016, leung_butterfly, gyongyosi2017entanglement, hahn_quantum_2019}, which result in long waiting times for the users of the network.

In contrast, there is a top-down approach to quantum networks, the so-called EBQNs \cite{pirker_modular_2018, pirker_quantum_2019, miguel-ramiro_optimized_2021, meignant_2019, gyongyosi2019opportunistic, miguel2020delocalized}, which prevent these waiting times and are the focus of this paper. In such networks, multipartite states are generated beforehand and stored until a request arrives. These resource states are then manipulated by local operations to establish the desired target state, e.g., a Bell pair shared between two nodes, without further using quantum communication. Thanks to the pre-generated entanglement, the time to achieve the target state only amounts to classical communication between devices. Thus, these resource states in EBQNs circumvent long waiting times at the cost of demanding network devices with long-time quantum memories. However, the users of the network have minimal functionality in contrast to the bottom-up approach. EBQNs have two main features that define the network. On the one hand, there is the network's entanglement topology, which is the entanglement structure of the pre-generated resource state. On the other hand, there is the actual physical structure of the network, where the nodes and quantum devices are placed. These two structures do not need to be the same and this introduces interesting aspects which are not present in the bottom-up approach. For instance, one can enable a direct entanglement link between two network devices that are not connected in the physical setting \cite{schoute2016shortcuts}. In \cite{pirker_quantum_2019} three phases in EBQNs are identified: 
\begin{enumerate}
    \item Dynamic phase: Entangled (resource) states are first distributed and established among the network nodes. In EBQNs, the distribution can be done before requests, when the network is idle, therefore avoiding the associated waiting times.
    \item Static phase: Once the resource state is established, the network stores it for future requests. Quantum memories used for said storage are experimentally challenging \cite{pirker_modular_2018}, and thus resource states that minimize the required storage are preferable.
    \item Adaptive phase: In this final phase the resource state is manipulated locally by the nodes. This is triggered by either the request of the users of the network or by a failure of devices in a quantum network. This last phase is the focus of this paper. 
\end{enumerate}

Additionally, in \cite{pirker_quantum_2019} a quantum network stack focused on structures and architectures of EBQNs following the design of the classical network stack (Open Systems Interconnection model \cite{osi_1980}) is proposed. This allows one to divide the design and analysis of EBQNs into hierarchical layers. In \cite{miguel-ramiro_optimized_2021}, the top-down approach to quantum networks is further improved using the inherent structure’s flexibility. In particular, the entanglement topology is optimized to the desired functionality of the network, leading to minimal memory requirements of the resource states. This enhancement is only possible in this framework because as mentioned earlier, the bottom-up approach is limited by the underlying physical setting of the network.

Resource states are generally considered to be graph states \cite{hein_multiparty_2004, hein_entanglement_2006}, as they consist of a large class of highly entangled states which can be converted by local operations and classical communication to other entangled states (see, e.g., \cite{mannalath, fischer_2021, hahn_limitations}). Moreover, they can generate any state on a subsystem of reduced size via measurements only \cite{raussendorf_measurement, raussendorf_one}. However, deciding whether a given graph state can be transformed into a set of Bell pairs using only local operations and classical communication is NP-complete \cite{dahlberg_transforming}.

\subsection{Stabilizer formalism and graph states}\label{ssec:graph:states}

\subsubsection{Stabilizer formalism and local Cliffords}
The stabilizer formalism \cite{GottesmanThesis} is a compact way to describe quantum states efficiently, making use of the stabilizer group $\mathcal{S}$. The latter is a subgroup of the $N$-qubit Pauli group, $\mathcal{P}_N=\{\pm1,\pm i\}\times\{\mathds{1}, \sigma_x, \sigma_y, \sigma_z\}^{\otimes N}$, which does not include the element $-\mathds{1}$. Each element of the stabilizer group is a stabilizer operator and one can find the subset of those which are maximally independent, the so-called stabilizer generators. Such that, any element of $\mathcal{S}$ can be generated by the product of the stabilizer generators. Finally, a stabilizer state (a quantum state in the stabilizer formalism), for a given stabilizer group $\mathcal{S}$, can be defined as a simultaneous eigenstate with eigenvalue +1 of all the stabilizer generators of $\mathcal{S}$. This description requires only $N$ generators, in contrast to $2^N$ complex coefficients using a basis representation of the state vector. 

Consider a stabilizer state $|{\psi}\rangle$, defined by the stabilizer group $\mathcal{S}=\langle\{g_i\}\rangle$, where $g_i$ are the generators. Then a local Clifford (LC) operation $U$, which is defined as a unitary quantum operation that maps stabilizer states to stabilizer states \cite{nest_invariants_2005}, acting on $|{\psi}\rangle$ is $U|{\psi}\rangle=U g_i|{\psi}\rangle=U g_iU^{\dagger}U|{\psi}\rangle=g_i' U|{\psi}\rangle$, such that the state $U|{\psi}\rangle$ is stabilized by all $g_i'$. Note that quantum states that can be described using this formalism are a subset of all quantum states. Nevertheless, stabilizer states are widely used in the quantum information framework, as they include many different kinds of entanglement. 

\subsubsection{Graph states and transformations}
In the framework of EBQNs, graph states \cite{hein_multiparty_2004, hein_entanglement_2006} are used to describe resource states. They are a subclass of multi-qubit highly entangled states that can be represented as graphs $G = (V, E)$, where $V$ denotes a finite set containing the vertices and $E$ is a set whose elements are the edges between two vertices. Graph states are an instance of stabilizer states, such that the state associated with this graph $G$ corresponds to the unique $+1$ eigenstate of the stabilizers $K_{a}=\sigma^{(a)}_x\prod_{(a, b)\in E} \sigma^{(b)}_z$ for all $a\in V$. Equivalently, graph states can be described by a controlled-Z gate, $CZ = diag (1, 1, 1, -1)$, acting between any two qubits that are connected by an edge, i.e, $|G\rangle =\prod_{(a,b)\in E}CZ^{(a,b)}|+\rangle^{\otimes V}$, where $|+\rangle = \frac{1}{\sqrt{2}}(|0\rangle + |1\rangle)$ is the +1 eigenstate of $\sigma_x$. Moreover, Bell pairs, two-qubit maximally entangled states, are local-unitary-equivalent (LU-equivalent) to 2-qubit graph states, for instance, $|{B}\rangle = \frac{1}{\sqrt{2}}(|{0}\rangle |{+}\rangle + |{1}\rangle |{-}\rangle)$. The most natural extension of these maximally entangled states to $N$-qubit systems are $\text{GHZ}_N$ states, in the graph state basis they are defined as $|{\text{GHZ}_N}\rangle = \frac{1}{\sqrt{2}}\left(|{0}\rangle |{+}\rangle^{\otimes N-1} + |{1}\rangle |{-}\rangle^{\otimes N-1}\right)$. These represent one particular type of graph state where there is a central or \textit{root} qubit with $N-1$ edges connecting to all the other qubits, called \textit{leaf} qubits. We also consider graph states corresponding to one-, two-, and $k$-dimensional lattices as well as trees in this paper.

Graph states can be manipulated and transformed by certain quantum operations \cite{hein_multiparty_2004, hein_entanglement_2006} to other graph states. Throughout the text we make use of the following ones, which are described in detail in \cite{hein_entanglement_2006, mor_noisy}:
\newline
\textit{Local complementation:} Given some vertex of the graph, this operation inverts the edges connecting the neighbors of said vertex. Two graph states are said to be LC-equivalent if the corresponding graphs are related by a sequence of local complementations.
\newline
\textit{Local Pauli $Z$ measurement:} A vertex can be removed by applying a Pauli $Z$ measurement, up to local correction operations, as depicted in Fig.~\ref{fig:manipulation}.
\newline
\textit{Local Pauli $Y$ measurement:} This operation produces a local complementation and deletion of the measured vertex, up to local correction operations. It is graphically shown in Fig.~\ref{fig:manipulation}.
\newline
\textit{Local Pauli $X$ measurement:} This measurement acts as a local complementation on a neighboring qubit of the measured one, then local complementation on the measured qubit, followed by the deletion of it, and lastly repeating the local complementation on the neighboring qubit of step one.
\newline
\textit{Merging operation:} Two graph vertices can be merged into a single one, which has both of their neighborhoods. This operation is done via an entangling operation and a vertex deletion.
\newline
\textit{Full merging operation:} Corresponds to a merging operation in which the remaining qubit is also measured in the $Y$ basis. It is equivalent to a measurement in the Bell basis.

\subsection{Noisy stabilizer formalism}
In \cite{mor_noisy}, the NSF was introduced as a method to describe the manipulation of noisy graph states, which scales linearly in the number of qubits of the initial state, but exponentially in the size of the target state. The main idea behind this formalism is that states and noise operators are updated independently, which is in contrast with the usual treatment that updates the initial mixed state. The individual update is done using commutation relations between Pauli noise and local Cliffords describing manipulation operators, these relations are named \textit{update rules}. Additionally, the computation of these rules is simplified using the fact that any Pauli noise map acting on a graph state can be rewritten as a noise map with noise operators that only contain products of $Z$ and $\mathds{1}$ \cite{dur_stability_2004}, due to the stabilizer structure of graph states. Consider a manipulation operator $O$ acting on a noisy graph state, where one of the noise operators is $N$. Then $ON|G\rangle$  is rewritten as $\tilde{N}O|G\rangle$ using the update rules, where $\tilde{N}$ only contains $Z$ and $\mathds{1}$ and denotes the updated noise operator acting on the noise-less manipulated graph state, $O|G\rangle$. In general, one has a set of manipulations (e.g., Pauli measurements, local complementations) acting on a noisy graph state. Consider a state $\rho=|G\rangle\langle G|$ subject to $l$ Pauli-diagonal noise maps, $\mathcal{E}_l\dots \mathcal{E}_1\rho$, which is manipulated by $k$ different operations acting on a set of qubits ($\{b_i\}$). The final state is $O_k^{(b_k)}\cdots \, O_1^{(b_1)}\left(\mathcal{E}_l\cdots \mathcal{E}_1\rho\right){O_1^{\dagger}}^{(b_1)}\cdots \, {O_k^{\dagger}}^{(b_k)}$, which using the NSF can be rewritten as 
\begin{equation}\label{eq:nsf}
    \widetilde{\mathcal{E}}_l\cdots \widetilde{\mathcal{E}}_1\left(O_k^{(b_k)}\cdots \, O_1^{(b_1)}\rho \, {O_1^{\dagger}}^{(b_1)}\cdots \, {O_k^{\dagger}}^{(b_k)}\right),
\end{equation}
where $O_k^{(b_k)}\cdots \,O_1^{(b_1)}\rho \,{O_1^{\dagger}}^{(b_1)}\cdots\, {O_k^{\dagger}}^{(b_k)}$ is the noise-less manipulated state, and $\widetilde{\mathcal{E}}_i$ are the updated noise maps. These final noise maps are computed by updating each of their noise operators using the update rules for each manipulation operator $O_i^{(b_i)}$. In Eq.~\eqref{eq:nsf} one can see that the updated noise maps act on the updated graph state or target state, and thus, the size of their noise operators depends on the size of the target state. One can then apply the effective noise maps on the target state of reduced size, thereby avoiding the treatment of large density matrices as long as the target state is small. Notice that the NSF allows for an analytic treatment of noise and a complete description of noise propagation in graph state manipulation. The use of this analytic tool is in contrast with the numeric methods commonly used in the framework of quantum communication to simulate noise, such as Monte Carlo simulation \cite{coopmans_netsquid_2021, matsuo2018, Satoh2016}.

\section{Setting}\label{sec:setting}
\noindent
This work is in the framework of EBQNs, which we have described in Sec.~\ref{ssec:EBQN}. Our focus is on the adaptive phase, the last of three phases. During the two previous ones (dynamic and static) an $N$-qubit resource state is distributed and stored for a certain time in the nodes of the network. Then in the adaptive phase, this resource state is manipulated locally (using the quantum operations described in Sec.~\ref{ssec:graph:states}) to achieve a certain target state, e.g., a Bell pair or a $\text{GHZ}_3$ state, between some of the parties specified in a network request. So far, the study of the adaptive phase does not include the influence of imperfections in the networks arising from the two previous phases and the execution of resource state manipulation in the adaptive phase. Hence, in this paper, we consider a noisy scenario, which is further discussed in this section together with an introduction and motivation of the sets of resource states and target states that are used throughout the paper.

\subsection{Noise model}
To study the influence of noise, decoherence, and imperfections in the adaptive phase, we investigate the manipulation of noisy states. We consider that the local preparation of resource states can be done with high fidelity, and we study the influence of distributing states through noisy channels, storing the entangled qubits in quantum memories until they are needed, and manipulating them using imperfect local operations. All these processes are modeled by local depolarizing noise that acts on each of the qubits of the resource state independently followed by perfect manipulations. The local depolarizing noise model is defined as
\begin{equation}
    \mathcal{E}_a\rho = p\rho +\frac{1-p}{4}\sum_{i=0}^3\sigma_i^{(a)}\rho\sigma_i^{(a)},
\end{equation}
where $\sigma_0=\mathds{1}$, $\sigma_1=\sigma_x$, $\sigma_2=\sigma_y$ and $\sigma_3=\sigma_z$, $\rho=|\Psi\rangle\langle\Psi|$ denotes the noiseless resource state and $p$ is the probability that the state remains unchanged, while $1-p$ is the probability that qubit $a$ is depolarized, meaning that it has the completely mixed state $\mathds{1}/2$. Thus, the noisy resource state is $\mathcal{E}_1 \mathcal{E}_2 \cdots \mathcal{E}_N |\Psi\rangle\langle\Psi|$, and this one is then manipulated using local measurements or merging processes, such that it is transformed into a noisy version of the desired target state. 

Importantly, the use of a single-qubit channel is justified, as shown in \cite{wallnoefer_meas}, for the resource state generation via entanglement purification. Moreover, it also describes channel noise when transmitting qubits of a locally prepared resource state to remote parties through separate channels, and it also accurately describes decoherence and noise from storage, as qubits are located at different parties. Finally, this noise model also includes imperfect local operations as depolarizing noise followed by a perfect local operation is an accurate and quite general error model for noisy local operations. These local operations include single-qubit measurements and merging operations, where the latter is always performed between qubits that are in the same node of the network. Furthermore, the single-qubit depolarizing noise model can be considered a worst-case model for single-qubit noise, as shown in \cite{dur2005standard}. The error propagation during the manipulation process is fully analyzed, as the impact of the initial local noise on the target state is analytically studied using the NSF.

\subsection{Possible resource and target states}
We consider a certain target state, e.g., a Bell pair, and we study the noise impact on an average target state, e.g., a Bell pair between any two nodes of the network. The main analyzed target state in this paper is a single Bell pair, which is useful for different ranges of applications or demands of a quantum network. In particular, a single Bell pair can be used as a quantum communication channel or as a resource for quantum key distribution \cite{ekert_1991}. We also consider a $\text{GHZ}_3$ state as a target state, which has many applications as a multiparty entangled state in several protocols in quantum communication and cryptography, e.g., secret sharing \cite{markham_graph, Hillery1999} and the quantum Byzantine agreement \cite{benor_2005}, and also in quantum metrology \cite{shettell_graph, RevModPhys.90.035005, giovannetti2011advances, T_th_2014}. 

For the resource states, we consider both sets of small graph states, i.e., Bell pairs and $\text{GHZ}_3$ states, and large graph states, e.g., one and two-dimensional clusters and $\text{GHZ}_N$ states. On the one hand, small states as resources are easy to purify and refresh after a certain storage time. On the other hand, one needs to store several of these states shared between different parties, which requires larger memory and the use of merging operations. Note that merging operations have a bigger impact on noise than local measurements. Large cluster states as resource states are the opposite, as they are harder to prepare, purify, and refresh. However, due to their entanglement structure, a single qubit per node in the network is sufficient, minimizing the storage requirements, and merging operations are not required. 

\begin{figure}[!t] 
    \centering
    \subfloat[]{\includegraphics[width=0.44\columnwidth]{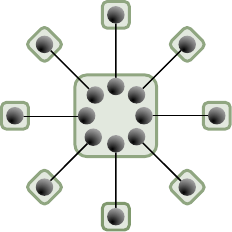}\label{fig:N:BP}}
    \hspace{0.9cm}
    \subfloat[]{\includegraphics[width=0.44\columnwidth]{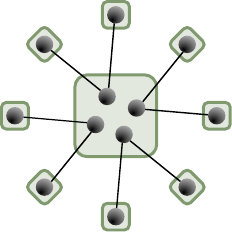}\label{fig:3GHZ}}
    \caption{Centralized network of eight external nodes (small squares) and a central node (big square). The circles are qubits and the solid lines represent entanglement between them. In (a) eight Bell pairs are distributed in the network and in (b) $\text{GHZ}_3$ states are distributed in the network, such that the leaf qubits are placed in the external nodes.}
    \label{fig:central}
\end{figure}

When studying sets of small states we consider a centralized network geometry where there is a central node and several external nodes as depicted in Fig.~\ref{fig:central} for the case with eight external nodes. Importantly, this switch-like structure has several qubits placed in the central node, and usually, merging-type manipulations are required in this node. 

As discussed in Sec.~\ref{ssec:EBQN}, EBQNs differentiate between the entanglement topology and the physical structure of the network. Thus, we can consider $k$-dimensional cluster states as the entanglement topology of the resource states. There are multiple ways to embed these states in a given physical geometry. This can be optimized with respect to the desired functionality of the network (possible target states) or the state generation. However, these two optimizations are distinct and might lead to contradictory results. In this paper, instead of optimizing over a certain target state, we consider average cases, where all possible target configurations are taken into account. For a single Bell pair as the target state, this means we consider the average geometrical distance between two nodes in the network. 

\section{Fidelity computation}\label{sec:fidelity}
\noindent
In this section, we describe how to compute the fidelity of a noisy target Bell pair using the NSF (see Sec.~\ref{sec:background}). Pauli measurements need to be done in a certain order, as correction operations depending on the outcome need to be applied on neighboring qubits. While this is irrelevant in a noiseless setting, it is crucial when considering noise. Notice that the noise maps from the target qubits and the manipulated ones need to be considered as both act on the target \cite{dur_stability_2004, hein_entanglement_2005}.

Due to the small size of a Bell pair as a target, the NSF allows us to derive an exact expression of the target state's fidelity regardless of the size of the resource state. This is possible since the NSF updates the noise maps of all the involved qubits in the manipulation. The resulting (reduced) state is of small size (two qubits) and the noise maps are of the same size, such that they can be applied and the noisy target state is efficiently computed. The number of noise maps is the same as in the initial state, i.e., linear in $N$ for the local noise model we consider, where $N$ is the size of the resource state. This is in contrast to the standard approach, where the initial noise is applied to the resource state using density matrices of the size $2^N\times 2^N$. Then this (large) quantum state is manipulated by applying matrices of the same size to it, which is extremely costly in terms of computational memory if the resource state is large (more than a few tens of qubits). Therefore, the manipulation of noisy graph states is in general a numerically hard problem, which can be efficiently solved by using the NSF for small target states.

In \cite{mor_noisy}, the case of a Bell pair as a target state where the resource state is subject to single-qubit depolarizing noise is presented, which coincides with the scenario considered in this paper. Importantly, the NSF makes use of the fact that any Pauli noise acting on a graph state can be written as a product of Pauli $Z$ operators and $\mathds{1}$ \cite{dur_stability_2004}. Then the results in \cite{mor_noisy} state that the final noise maps of the qubits measured in a Pauli basis can only take three forms 
\begin{align}
    \mathcal{M}_{\alpha}\rho'& = p\rho' + \frac{1-p}{2}\left(\rho' + Z_a Z_b\rho' Z_a Z_b\right), \label{eq:ma} \\
    \mathcal{M}_{\beta}\rho'& = p\rho' + \frac{1-p}{2}\left(\rho' + Z_a\rho' Z_a\right), \label{eq:mb}\\
    \mathcal{M}_{\gamma}\rho'& = p\rho' + \frac{1-p}{2}\left(\rho' + Z_b\rho' Z_b\right), \label{eq:mc}
\end{align}
where $a$ and $b$ are the labels of the target qubits that form the final Bell pair and $\rho'$ denotes the noiseless target state, $\rho' = |B\rangle \langle B|$. Moreover, the noise maps of both the target qubits, $a$ and $b$, and the qubits involved in a full-merging operation have the following form
\begin{equation}\label{eq:target:BP}
\begin{aligned}
    \mathcal{M}_{\text{target}}\rho' = p\rho' + \frac{1-p}{4} (\rho' + Z_a\rho' Z_a &+ Z_b\rho' Z_b \\ &+ Z_a Z_b\rho' Z_a Z_b ).
\end{aligned}
\end{equation} 
Thus, the final noisy state is $\mathcal{M}_{\alpha} \cdots \mathcal{M}_{\beta}\cdots \mathcal{M}_{\gamma}\cdots \mathcal{M}_{\text{target}}\rho'$, where each kind of map is applied a certain number of times, and the total number of maps corresponds to the total number of qubits involved in the manipulation together with the target qubits. Note that the form of these maps is such that applying $\mathcal{M}_i$ with parameter $p$ $x$ times is the same as applying it once with parameter $p^x$. Additionally, in \cite{mor_noisy} the so-called weight vector is defined, which describes the noise maps of the qubits measured in a Pauli basis. This vector is defined as $\boldsymbol{w} = (w_{\alpha}, w_{\beta}, w_{\gamma})$, where each component corresponds to the number of times each final noise map $M_{\alpha}$, $M_{\beta}$, or $M_{\gamma}$ is applied. Also, the number of full merging operations is defined as $t$. Thus, the number of local measurements required in the manipulation corresponds to $\sum_iw_i$ and $2t$ is the number of full-merged qubits.

Furthermore, using these updated noise maps and their properties an exact expression for the fidelity of the target Bell pair can be computed \cite{mor_noisy}, 
\begin{equation}\label{eq:general:fidelity:BP}
    F(p,\boldsymbol{w}, t)=\frac{1}{4}\left(1+p^{2+2t}\sum_{\substack{i,j\in \{\alpha, \beta, \gamma\} \\ i\neq j}}p^{w_i + w_j}\right).
\end{equation}
So the fidelity of the target state can be directly determined from $\boldsymbol{w}$ and $t$, such that it depends mainly on the number of operations and the order of consecutive measurements.

In the scenarios where the target state is a Bell pair, we compute the corresponding $\boldsymbol{w}$ and $t$ for an exact fidelity together with the approximation of the fidelity in the low error regime (LER), such that $1-p=\epsilon$ for small $\epsilon$. This fidelity is 
\begin{equation}\label{eq:general:fidelity:BP:approx}
    F(\epsilon, \boldsymbol{w}, t)\simeq  1-\frac{1}{2}(3 +3t + w_{\alpha} + w_{\beta} + w_{\gamma})\epsilon.
\end{equation}
Note that to compute the latter one just needs a counting argument on how many local Pauli measurements and full-merging operations are required. Moreover, Eq.~\eqref{eq:general:fidelity:BP:approx} makes clear that a full merging procedure has a higher impact on the fidelity than a single-qubit measurement. In the following, we make use of the function $g(x)=\frac{1-(-1)^x}{2}$ to compactly express weight vectors.

\section{Results for a single Bell pair as a target state}\label{sec:BP}
\noindent
In this section, we present the results of achieving a single Bell pair via the manipulation of different resource states. The target Bell pair is between two arbitrary qubits $a$ and $b$ and the fidelity is computed using the results of Sec.~\ref{sec:fidelity}. There are two different types of procedures: (i) the manipulation of small states via full-merging (Sec.~\ref{ssec:BP:BP} and \ref{ssec:BP:3GHZ}), and (ii) the manipulation of a large resource state via single-qubit measurements (Sec.~\ref{ssec:BP:NGHZ} - \ref{ssec:BP:tree}), such that $t=0$.

\subsection{Resource: Several Bell pairs}\label{ssec:BP:BP}
We assume a centralized network structure that has a central node and $N$ external nodes \cite{Cuquet2012, coopmans_netsquid_2021, vardoyan_stochastic_2019, Vardoyan2020, caleffi_2020, nain_2020, nain2022analysis}. Then, $N$ Bell pairs are distributed such that each pair is established between an external node and the central node, requiring $2N$ qubits. This switch structure is depicted in Fig.~\ref{fig:N:BP}. The performed manipulation to achieve a single Bell pair between any two external nodes is a full merging between the two qubits in the central node. Such that $\boldsymbol{w}=(0,0,0)$ and $t=1$, thus, the resulting fidelity in the LER is $F(\epsilon)\simeq 1-3\epsilon$.

\subsection{Resource: Several three-qubit GHZ states}\label{ssec:BP:3GHZ}
We consider several $\text{GHZ}_3$ states in the centralized network structure. The states are distributed such that the root qubits are placed in the central node and each leaf is in an external node, as shown in Fig.~\ref{fig:3GHZ}. There are $N/2$ $\text{GHZ}_3$ states that require $3N/2$ qubits, for even $N$. 

There are two possible scenarios to achieve a single Bell pair between any two external nodes. On the one hand, the two target qubits correspond to the same GHZ state. Therefore, the final fidelity is the same as the case of $N=3$ in Sec.~\ref{ssec:BP:NGHZ}, $F(\epsilon,N)\simeq1-2\epsilon$, i.e., larger than the fidelity obtained when fully merging two Bell pairs. On the other hand, the target qubits can correspond to leaves of different $\text{GHZ}_3$ states. In this case, one needs to $Z$-measure the extra leaves in the two $\text{GHZ}_3$ states and then fully merge the two remaining pairs. Such that $\boldsymbol{w}=(0,1,1)$ and $t=1$, thus, the fidelity in the LER is $F(\epsilon)\simeq 1-4\epsilon$.

\subsection{Resource: N-qubit GHZ state}\label{ssec:BP:NGHZ}
Consider a centralized structure with $N-1$ external nodes. An $\text{GHZ}_N$ state is distributed such that the root qubit is placed in the central node and the leaves in the external ones, which requires a total of $N$ qubits. The performed manipulation to achieve a single Bell pair between any two leaf qubits is the $Z$-measurement of all the other leaf qubits, followed by the $Y$-measurement of the root qubit. Such that $\boldsymbol{w}=(N-2,0,0)$ and $t=0$, thus, the fidelity in the LER is $F(\epsilon,N)\simeq1-\frac{1}{2}(N+1)\epsilon$. Note that the fidelity decreases as the size of the GHZ state increases. Thus, a $\text{GHZ}_3$ state leads to a Bell pair with the highest fidelity.

\subsection{Resource: 1D cluster}\label{ssec:BP:1D}
Consider an $N$-qubit 1D cluster with periodic boundary conditions (PBCs). Each cluster
qubit is placed in a node of the network, each with two
neighbors.

In the manipulation to achieve a Bell pair between two arbitrary qubits, there are two kinds of qubits that are of importance. On the one hand, there are the qubits between the pair, which define the shortest path, we refer to these as \textit{inner neighbors}. On the other hand, there are the neighboring qubits surrounding the path, which we refer to as \textit{outer neighbors}. The number of inner neighbors is $n_{1\text{D}}$, such that $0\leq n_{1\text{D}} \leq [N-g(N)]/2-1$, where $g(x)$ is defined in Table~\ref{tab:notation}, and the number of outer neighbors is $m_{1\text{D}}=2$. The manipulation performed is a $Y$ measurement of inner neighbors and a $Z$ measurement of outer neighbors, as shown in Fig.~\ref{fig:BP:1D}.

The noise that affects the final Bell pair is the one from the target qubits and the inner and outer neighbors, as stated in \cite{dur_stability_2004, hein_entanglement_2005}. Furthermore, as seen in \cite{mor_noisy}, the order in which the inner neighbors are measured is relevant. Here we restrict to a strategy where qubits are measured sequentially along the path, called side-to-side in \cite{mor_noisy}. The different strategies vary slightly in the achievable fidelity, with a relative difference of up to $2.5\%$ in the regime where the fidelity of the resulting state is $F>1/2$, i.e., is still entangled. Following the same calculation as outlined in \cite{mor_noisy}, but taking PBCs and hence two extra outer neighbors into account, one can compute the resulting fidelity. The weight vector is given by 
\begin{equation}\label{eq:1D:side:to:side}
    \boldsymbol{w}_{1\text{D}}=\left(\frac{n_{1\text{D}}+g(n_{1\text{D}})}{2},\, 1, \, \frac{n_{1\text{D}}-g(n_{1\text{D}})}{2}+1\right),
\end{equation}
which results in a LER fidelity of $F(\epsilon, n_{1\text{D}}) \simeq 1-\frac{1}{2}(n_{1\text{D}}+5)\epsilon$.

\subsection{Resource: 2D cluster}\label{ssec:BP:2D}
Consider an $N$-qubit 2D cluster with PBCs. Each cluster qubit is placed in a node of the network, each with four neighbors. We analyze two methods to achieve the target pair:

\subsubsection{Y method}
\begin{figure*}[!t]
    \centering
    \begin{minipage}{\columnwidth}
    \subfloat[]{\includegraphics[width=0.98\columnwidth]{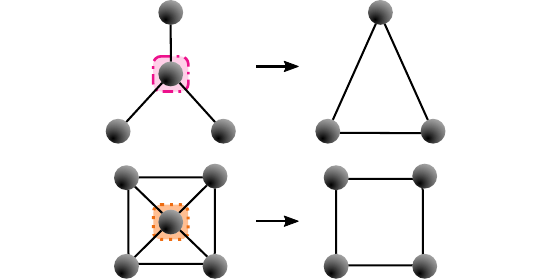}\label{fig:manipulation}}
    \vspace{1.5cm}
    \subfloat[]{\includegraphics[width=\columnwidth]{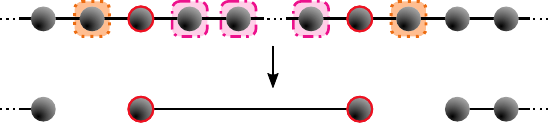}\label{fig:BP:1D}}
    \end{minipage}
    \hfill
    \begin{minipage}{0.98\columnwidth}
    \subfloat[]{\includegraphics[width=\columnwidth]{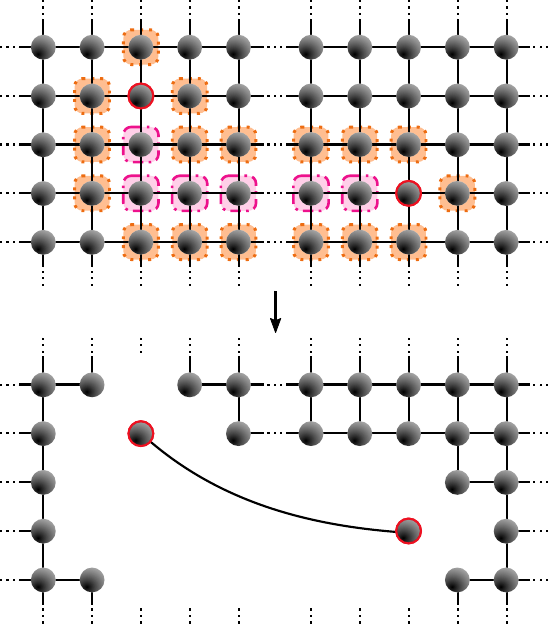}\label{fig:BP:2D}}
    \end{minipage}
    \subfloat{\includegraphics[width=0.98\columnwidth]{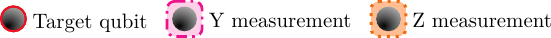}}
    \caption{The dark circles are qubits, the solid lines represent the entanglement between qubits and the dotted lines represent the extension of the cluster following the presented structure. Thus, only part of a large cluster is explicitly drawn in (b) and (c). (a) is the graphical representation of the local Pauli $Y$ and $Z$ measurements in graph states. (b) is the graphical representation of the manipulation of a 1D cluster with PBCs into a Bell pair. (c) is the graphical representation of the manipulation of a 2D cluster with PBCs into a Bell pair using the $Y$ method. At the top of (b) and (c), the initial situation is depicted, where the target qubits have been chosen and the corresponding manipulation is established. At the bottom of (b) and (c), the manipulation has been performed and the resulting Bell pair and the remaining graph state are presented.}
\end{figure*}

This method consists in deciding the shortest path between the two target qubits followed by a $Z$ measurement of the outer neighbors and a $Y$ measurement of the inner neighbors, as shown in Fig.~\ref{fig:BP:2D}. 
Note that the shortest path is not unique, one can choose a stairway-like path, which consists of several lines connected by corners, two straight lines connected by a corner, or in some cases a single straight line. Importantly, a corner in the path reduces the number of outer neighbors by one. Thus, fewer measurements need to be performed and fewer noise maps affect the final Bell pair, leading to higher fidelity overall. However, analyzing the noise of a path that has one or several corners is highly dependent on the position of the targets and the corners. Thus, a general result for a path that includes corners is rather hard. Despite that, if a particular geometrical situation is given, one can find the corresponding weight vector and thus, the exact fidelity. Here, we give counting arguments for the most general case, where the path has one corner, and the case where the path is a stairway. Moreover, we focus on the lower-bound approximation that any two qubits in a 2D cluster can be connected via a straight line, as a straight-line path has more outer neighbors, thus leading to a lower fidelity of the final Bell pair. 

We can use the results presented in Sec.~\ref{ssec:BP:1D} to compute the fidelity for a straight path. Consider $n_{2\text{D}}$ inner neighbors, such that $0\leq n_{2\text{D}} \leq \sqrt{N}-1-g(N)$, and $m_{2\text{D}}=2n_{2\text{D}}+6$ outer neighbors, as each inner neighbor has two outer neighbors and the target qubits have three each. Using the NSF, one can see that the noise maps of the outer neighbors of the inner qubits will have the same form as the noise maps of the corresponding inner qubit. Therefore, we can use the results presented in Sec.~\ref{ssec:BP:1D} and generalize the weight vector for a 2D cluster, such that for this resource state the new weight vector is $\boldsymbol{w}_{2\text{D}}^{(Y)}=3\boldsymbol{w}_{1\text{D}}$, where $\boldsymbol{w}_{1\text{D}}$ corresponds to the weight vector for the 1D cluster (for any order of measurements) using $n_{2\text{D}}$ instead of $n_{1\text{D}}$. Note that the values that $n_{2\text{D}}$ and $n_{1\text{D}}$ can take are different, in particular on average $n_{2\text{D}}$ is much smaller than $n_{1\text{D}}$, as $n_{2\text{D}} \propto\sqrt{N}$ while $n_{1\text{D}} \propto N$. The fidelity in the LER is $F(\epsilon, n_{2\text{D}})\simeq 1-\frac{1}{2}(3n_{2\text{D}}+9)\epsilon$. Each corner in the path reduces the number of outer neighbors by one, such that a stairway path has the minimum number of outer neighbors. The fidelity in the LER hence reduces to $F(\epsilon, n_{2\text{D}})\simeq 1-\frac{1}{2}(3n_{2\text{D}}+8)\epsilon$ for one corner, and 
\begin{equation}\label{eq:fid:BP:2D:Y}
    F(\epsilon, n_{2\text{D}})\simeq 1-\frac{1}{2}(2n_{2\text{D}}+9)\epsilon
\end{equation}
for the stairway, which has $n_{2\text{D}}$ inner neighbors and $n_{2\text{D}}+6$ outer neighbors. Note that only for specific configurations of target nodes a stairway connection is possible. 

\subsubsection{X method}
In \cite{hahn_quantum_2019} an alternative method to get a Bell pair is presented. We refer to the alternative method as \textit{$X$ method} because as a first instance, one decides on the shortest path between two qubits and $X$-measures the inner neighbors consecutively. Next, in the resulting state, all the neighboring qubits of the target qubits are $Z$-measured.

The $X$ method applied in a 2D cluster is optimal if the chosen path has many corners, as proved in \cite{mannalath}. Moreover, that requires fewer measurements than the $Y$ method. In particular, if an outer neighbor is neighboring two inner neighbors, after the $X$ measurements, that qubit is no longer connected to the target qubits. For a stairway path of $n_{2\text{D}}$ inner neighbors, the number of qubits that have to be $Z$-measured is fixed to six, as shown in Appendix~\ref{a:x:protocol}. The associated weight vector of a Bell pair achieved via the $X$ method is
\begin{equation}\label{eq:x:protocol}
    \boldsymbol{w}_{2\text{D}}^{(X)} = \left(0, \, \frac{n_{2\text{D}}-g(n_{2\text{D}})}{2}+3, \, \frac{n_{2\text{D}}+g(n_{2\text{D}})}{2}+3 \right).
\end{equation}
Then, the fidelity in the LER is
\begin{equation}\label{eq:fid:BP:2D:X}
    F(\epsilon, n_{2\text{D}}) \simeq 1 - \frac{1}{2}(n_{2\text{D}} + 9)\epsilon.
\end{equation}

\subsection{Resource: kD cluster}\label{ssec:BP:kD}
Consider an $N$-qubit $k$-dimensional cluster with PBCs. Each cluster qubit is placed in a node of the network, each with $2k$ neighbors. In each direction, the size of the cluster is $\sqrt[k]{N}$.

The performed manipulation follows the $Y$ method, presented in Sec.~\ref{ssec:BP:2D}. The number of inner neighbors, $n_{k\text{D}}$, is such that $0\leq n_{k\text{D}}\leq k[\sqrt[k]{N}-g(N)]/2-1$, and the number of outer neighbors is $m_{k\text{D}}=2(k-1)n_{k\text{D}}+2(2k-1)$ when considering the lower-bound approximation that the shortest path is a straight line, as in Sec.~\ref{ssec:BP:2D}. Moreover, following the 2D case, we can further generalize the weight vector, such that $\boldsymbol{w}_{k\text{D}} = (2k-1) \boldsymbol{w}_{1\text{D}}$, where $\boldsymbol{w}_{1\text{D}}$ corresponds to the weight vector for the 1D cluster (for any order of measurements) using $n_{k\text{D}}$. Note that the values of $n_{k\text{D}}$ are different for each $k$ and in average $n_{k\text{D}}$ decreases as $k$ increases. The fidelity in the LER is $F(\epsilon, k, n_{k\text{D}})\simeq 1-\frac{1}{2}[(n_{k\text{D}}+2)(2k-1)+3]\epsilon$.

\subsection{Resource: Tree cluster}\label{ssec:BP:tree}
A tree cluster is a graph state that can be characterized by a \textit{branching vector}, $\boldsymbol{b}=(b_1,b_2,\dots, b_d)$ and $d$ denotes the depth of the tree cluster. In particular, we study a binary tree, which means that $b_i = 2, \, \forall i$, such that the total number of qubits is $N = 2^{d+1}-1$. Each qubit of the cluster is placed in a node of the network. The performed manipulation to achieve a Bell pair follows the $Y$ method. Note that how much of the tree state is measured out depends on the target qubits. We consider the worst case, the one that destroys most of the tree, which is when the two target qubits are in different branches of the tree. Moreover, we assume that the two target qubits are not at the bottom of the tree and that the path between them has $n_{\text{tree}}$ inner neighbors, such that $n_{\text{tree}} \propto \log_2 N$. There are $m_{\text{tree}}=n_{\text{tree}}+3$ outer neighbors, as each inner neighbor has an outer neighbor but the root qubit of the tree, which has no outer neighbor and each of the target qubits has two outer neighbors. If we take the lower-bound approximation that the root qubit has an outer neighbor, then the corresponding weight vector is $\boldsymbol{w}_{\text{tree}} = 2\boldsymbol{w}_{1\text{D}}$, where $\boldsymbol{w}_{1\text{D}}$ corresponds to the weight vector for the 1D cluster (for any order of measurements) using $n_{\text{tree}}$. The fidelity in the LER is $F(\epsilon, n_{\text{tree}}) \simeq 1-(n_{\text{tree}}+4)\epsilon$.

\subsubsection{Increasing the width of the tree} 
We are now interested in structures such that $b_1>2$. Consider $\boldsymbol{b}=(3, 2, \dots, 2)$, then $N=3 \times 2^d-2$. We again assume that the two target qubits are not at the bottom of the tree and that the path between them has $n_{\text{tree}}$ inner neighbors. Each inner neighbor has an outer neighbor and the target qubits have two each, thus, $m_{\text{tree}}=n_{\text{tree}}+4$. The weight vector is $\boldsymbol{w}_{\text{tree}} = 2\boldsymbol{w}_{1\text{D}}$, as the lower-bound of the binary tree. In general, as we increase the width of a binary tree the number of outer neighbors of the root qubit increases. Such that $m_{\text{tree}}=n_{\text{tree}}+b_1+1$, for a certain $n_{\text{tree}}$, considering the worst-case scenario.

\section{Comparison of resources}\label{sec:comparison}
In Table~\ref{tab:BP} we give the fidelity in the LER for all the studied resource states.
\begin{table}[!t]
\caption{Fidelity in the LER for a Bell pair between any two nodes in a network of $N$ nodes depending on the resource state (derived in Sec.~\ref{sec:BP}).
\label{tab:BP}}
\centering
\begin{tabular}{|l|l|}
\hline
\textbf{Resource state} & \textbf{Fidelity in the LER} \\ \hline \hline
Several Bell pairs & $1-3\epsilon$ \\ 
Several $\text{GHZ}_3$ states & $1-4\epsilon$ \\ 
$\text{GHZ}_N$ state & $1-\frac{1}{2}(N+1)\epsilon$ \\ 
1D cluster & $1-\frac{1}{2}(n_{1\text{D}}+5)\epsilon$ \\ 
2D cluster: \textit{$Y$ method straight} & $1-\frac{1}{2}(3n_{2\text{D}}+9)\epsilon$ \\ 
2D cluster: \textit{$Y$ method stairway} & $1-\frac{1}{2}(2n_{2\text{D}}+9)\epsilon$ \\
2D cluster: \textit{$X$ method stairway} & $1 - \frac{1}{2}(n_{2\text{D}} + 9)\epsilon$ \\
$k$D cluster & $1-\frac{1}{2}[(n_{k\text{D}}+2)(2k-1)+3]\epsilon$ \\
Binary tree cluster & $1-(n_{\text{tree}}+4)\epsilon$ \\ \hline
\end{tabular}
\end{table}

\subsection{Sets of small graph states}
Here we compare the results of using several Bell pairs and several $\text{GHZ}_3$ states as resources in a switch-type structure, presented in Sec.~\ref{ssec:BP:BP} and Sec.~\ref{ssec:BP:3GHZ}. There are three cases: (i) several Bell pairs, (ii) several $\text{GHZ}_3$ when the two target qubits correspond to the same GHZ state, and (iii) several $\text{GHZ}_3$ states when the two target qubits do not correspond to the same GHZ state. Overall, (ii) is optimal as it has the highest fidelity. This is followed by (i) and last (iii). Therefore, we can say that a $\text{GHZ}_3$ state is a more robust resource than two Bell pairs. Additionally, we also consider the requirement of the number of qubits for a fixed size of the network, $N$ external nodes. Distributing Bell pairs requires $2N$ qubits whereas $\text{GHZ}_3$ states require $3N/2$. Thus, considering multiparty entangled states, i.e., $\text{GHZ}_3$ states, is better in terms of memory.

\subsection{Y method vs. X method}
In Sec.~\ref{ssec:BP:2D}, we have analyzed two methods to achieve a single Bell pair, the $Y$ method and the $X$ method. Both methods are optimal for a stairway path, and thus, we are going to compare these cases. In terms of measurements, for a fixed number of inner neighbors $n_{2\text{D}}$, the $X$ method requires $n_{2\text{D}}+6$ measurements, whereas the $Y$ method requires $2n_{2\text{D}}+6$. This directly affects the fidelity, as $F_X-F_Y = n_{2\text{D}}\epsilon$, where $F_Y$ and $F_X$ correspond to Eqs.~\eqref{eq:fid:BP:2D:Y} and \eqref{eq:fid:BP:2D:X}. Thus the $X$ method has a higher fidelity for any $n_{2\text{D}}$.

\subsection{Optimal dimension of cluster} \label{sssec:BP:optimal:k}
\begin{figure}[!t]
    \centering
    \subfloat[]{\includegraphics[width=\columnwidth]{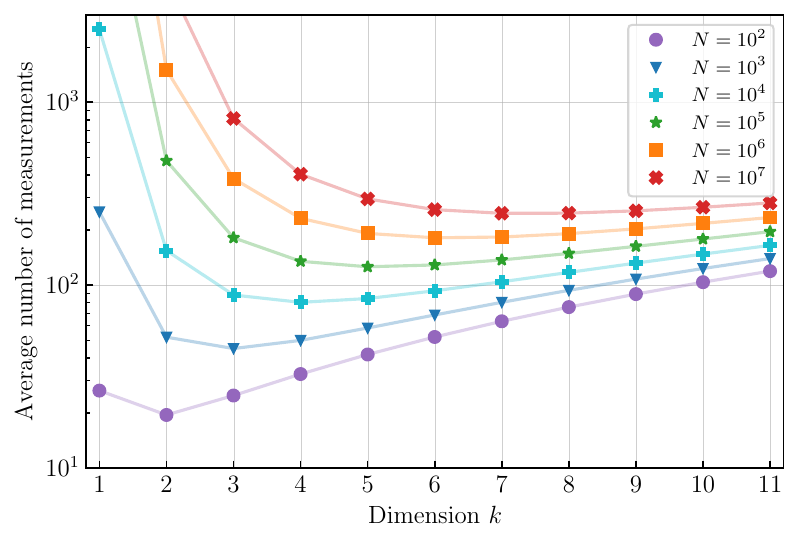}\label{fig:optimal:dimension}}
    \vfill
    \subfloat[]{\includegraphics[width=\columnwidth]{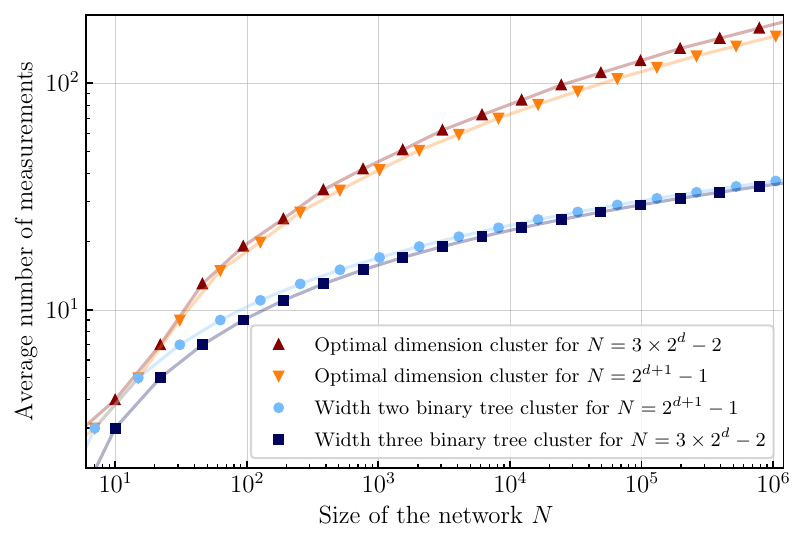}\label{fig:tree_vs_k}}
    \caption{(a) Log plot of the average number of measurements required to achieve a Bell pair from a $k$-dimensional cluster, in terms of the dimension of the cluster. For very large systems ($N\sim 10^5$), the value for $k=1$ is not included in the plot as it is too large. (b) Log-log plot of the average number of measurements in terms of the size of the size $N$ of the network. The different series correspond to different resource states. For $N=3\times 2^{d}-2$, the width-three binary cluster and the optimal dimension cluster marked with triangles are presented. For $N=2^{d+1}-1$, the width-two binary cluster and the optimal dimension cluster marked with inverted triangles are presented. Note that the values in (a) and the values of the two optimal cluster series in (b) are determined using Eq.~\eqref{eq:optimal:cluster} and ignoring the boundary effects due to the restriction to integer values of $\sqrt[k]{N}$.}
\end{figure}

We consider networks of size $N$, which implies that for $k$-dimensional clusters, the length in each dimension $n_{k\text{D}} \propto \sqrt[k]{N}$. Hence the (average) distance between any two nodes in the network decreases significantly with $k$, while the number of outer neighbors among the path increases (linearly). As shown in Sec.~\ref{ssec:BP:kD}, this leads to a different number of required measurements, and thus different fidelities. Our aim is to minimize the number of measurements, which is directly linked to the dimension and the size of the cluster. 
The average number of inner qubits of the path is $\bar{n}_{k\text{D}}(k, N) = \frac{k}{4}[\sqrt[k]{N}-g(N)]-\frac{1}{2}$, where we use that the average distance among each dimension is $\sqrt[k]{N}/4$ (the worst case would just by larger by a factor of $2$ due to PBCs). This leads to a required averaged number of measurements for the manipulation
\begin{equation}\label{eq:optimal:cluster}
    \bar{n}_{k\text{D}}+\bar{m}_{k\text{D}} = \frac{2k-1}{2}\left[\frac{k}{2}\left(\sqrt[k]{N}-g(N)\right)+3\right].
\end{equation}
Importantly, as the dimension of the cluster increases the connectivity of each qubit increases, whereas the average distance of the path decreases. Thus, for a fixed size of the cluster $N$, the average optimal dimension for a minimum number of measurements, $k'$, can be found. In Fig.~\ref{fig:optimal:dimension}, the required averaged number of measurements is shown in terms of the dimension of an $N$-qubit cluster. From the results of Fig.~\ref{fig:optimal:dimension} we can see that the minimum number of measurements corresponds to a certain dimension of the cluster which increases as the size ($N$) increases. For example, take $N=10^2$, then the optimal dimension is $k'=2$, such that $\bar{n}_{k\text{D}}=5$ and $\bar{m}_{k\text{D}}=15$ are minimal. Notice the moderate increase in the required number of measurements with the total size of the network $N$.  

\subsection{Tree cluster vs. optimal dimension cluster}\label{sssec:BP:tree:vs:cluster}
To compare tree graphs and optimal-dimensional clusters, we focus on the required number of measurements in the manipulation, as we have seen above the number of measurements directly impacts the fidelity of the final Bell pair. For the binary tree, $2n_{\text{tree}}-1$ measurements are required, which in average it is $2\log_2(N+1)-3$. Note that if the width of the tree is increased and $N$ is fixed, the overall required measurements are the same, but the average is smaller, the latter being $2\log_2((N+2)/3)-1$. This is a lower average than the pure binary tree, as one can see in Fig.~\ref{fig:tree_vs_k}. Moreover, the optimal-dimensional cluster requires more measurements than a tree cluster, as shown in Fig.~\ref{fig:tree_vs_k}. However, when manipulating a tree cluster, we destroy most of the entanglement, leaving a lot of disconnected small clusters. Whereas, after the manipulation of an optimal-dimensional cluster, most of the state remains entangled and can be further used to establish more connections. Therefore, we can say that a tree cluster is a better resource state for a one-shot scenario than the optimal-dimensional cluster in terms of achievable fidelity. 

\subsection{Thresholds for the error probability}
\begin{figure}[!t]
    \centering
    \includegraphics[width=\columnwidth]{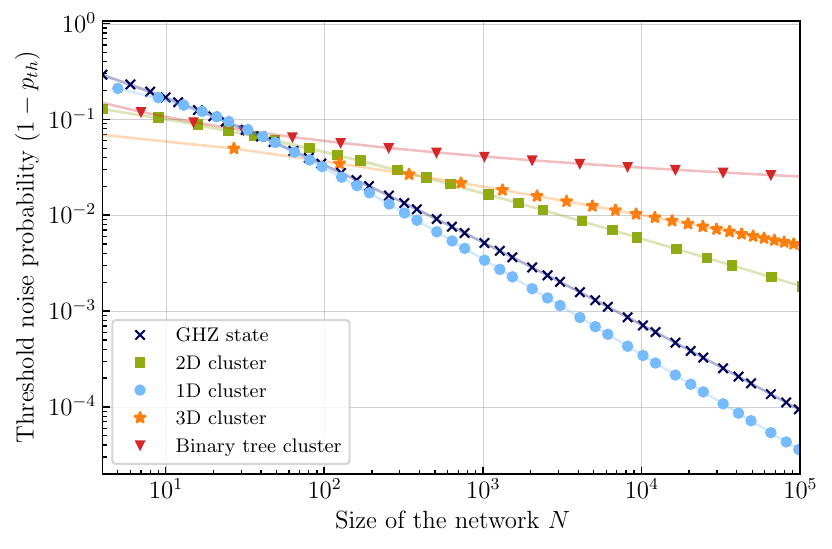}
    \caption{Log-log plot of the threshold value for the single-qubit noise probability for producing an entangled Bell pair, as a function of the total size $N$ of the network. Different curves correspond to different resource states, e.g., 1D, 2D, 3D and tree clusters and GHZ states. The worst-case scenario, i.e., the maximum possible distance between target nodes, is considered.}
    \label{fig:threshold}
\end{figure}

For the large resource states, the fidelity of a resulting Bell pair depends on the size of the cluster, such that if $N$ is too large the fidelity drops below $0.5$ and the resulting pair is no longer entangled. In Fig.~\ref{fig:threshold}, the threshold for the noise probability ($1-p_{th}$) for which the fidelity of the Bell pair is lower than $0.5$ is presented as a function of $N$, the size of the resource cluster. To compute the results presented in this section we use the exact expressions of the fidelity, which can be computed using the weight vectors and Eq.~\eqref{eq:general:fidelity:BP}, to get a more accurate result than using the fidelity in the LER. In contrast to the previous subsection, we consider a worst-case scenario here, i.e., the maximum possible distance between any two nodes, which implies that above the error threshold, one can produce an entangled Bell pair between any two nodes. The results in Fig.~\ref{fig:threshold} imply that large-scale GHZ states and 1D clusters are not suitable resources in an EBQN when taking noise and imperfections into account. However, 2D, 3D, and binary tree clusters have a good tolerance to noise in large-scale systems. Importantly, $1-p_{th}$ for these resource states does not go below $10^{-3}$ even for large ($\sim 10^5$) clusters. Notice that for large enough $N$, the 3D cluster is better than the 2D one, which is in agreement with the results of the optimal dimension cluster. Moreover, we see that the binary tree cluster has a better tolerance to noise than the 2D and 3D clusters, a behavior that we have already seen in the previous results. 

\section{Extension to a three-qubit GHZ state as a target state}\label{sec:3GHZ}
\noindent
We now generalize our results to a single $\text{GHZ}_3$ state as the target, where the state is shared between three arbitrary qubits $a$, $b$, and $c$. Then, one just needs to connect two of the qubits to the third one. Thus, from the three possible paths ($a$ to $b$, $b$ to $c$, $c$ to $a$) one chooses the two shortest ones and establishes the target state. Assume that the two shortest paths are $a$ to $b$ and $b$ to $c$, then $a$ and $c$ denote the leaf qubits of the target state and $b$ the root of the target state. If the two leaf qubits and their paths to the root qubit do not have joined neighbors, then the results derived for the Bell pair in Sec.~\ref{sec:BP} can be easily generalized for this case. To understand this, one can think of this $\text{GHZ}_3$ state as two Bell pairs that share one qubit, which results in being the root qubit. Then, under the consideration that the two leaf qubits and their paths to the root qubit do not have joined neighbors, these two pairs are essentially independent of each other. In Appendix~\ref{a:3GHZ}, we show that we can extend the weight vector formalism for this case and we derive an exact expression of the fidelity for the noisy $\text{GHZ}_3$ state. Moreover, we give the results for large cluster states manipulated by single-qubit measurements into a $\text{GHZ}_3$ state. The analysis of optimal-dimensional clusters is analogous to the one derived for the Bell pair as a target, taking into account that now there are two paths, one from $a$ to $b$ and another from $b$ to $c$. We remark that the scenario of transforming a 1D cluster into GHZ states has also been studied in \cite{jong_2022} for a system of up to a few tens of qubits.

\section{Conclusions and outlook}\label{sec:conclusion}
\noindent
In this article, we have investigated the influence of noise and imperfections in EBQNs. We found that there exist classes of resource states that allow one to produce Bell pairs and few-qubit GHZ states with moderate overhead, leading to high-fidelity entangled output states. In this case, it resulted that high-dimensional cluster states or tree graphs are favorable in terms of the required number of measurements, and hence in the achievable fidelity. Additionally, regarding storage requirements, these resource states are optimal as a single qubit per node is sufficient. Nevertheless, we have also considered resource states formed by a set of small states, i.e., Bell pairs or few-qubits GHZ states, which have higher storage requirements. These are beneficial because their entanglement can be entirely or partly easily refreshed. We conclude that a $\text{GHZ}_3$ state is a more robust resource than two Bell pairs.

However, in a real network one is typically not interested in single requests, but there might be another request after the first one was fulfilled. More generally, multiple requests that should be fulfilled in parallel are conceivable (see, e.g., \cite{bravyi}). In these cases, not only the achievable fidelity is relevant, but also the structure of the remaining entanglement in the resource state. GHZ states, 1D cluster states, and tree graphs are not favorable in this respect as most of the entanglement is destroyed during the preparation process of the Bell pair, and the remaining resource cannot be used directly to fulfill further requests. We remark that this is not true for higher dimensional clusters with $k\geq 2$, where multiple Bell pairs can be generated in parallel. Cutting 1D lines out of a multi-dimensional cluster leaves one with a connected cluster, and crossings can be avoided by using the third dimension. But even a 2D structure suffices, as an embedded butterfly state \cite{leung_butterfly, epping_robust_2016} can be used to achieve crossings.

Another interesting issue is concerned with the generation of multipartite target states from small elementary structures, e.g., the preparation of 1D or 2D cluster states from elementary Bell pairs or $\text{GHZ}_3$ states. An analysis of these processes can be performed; however, a direct assessment of the final state fidelity cannot be done using the NSF. While there exists an efficient description of resulting states and noise processes after merging in terms of stabilizers and noise operators acting on states, computation of fidelity involves an exponential overhead. One can however provide an indirect assessment of the quality of the resulting resource by computing average fidelities of Bell pairs produced from the cluster state. We note that recently a method to approximately determine the fidelity of cluster states efficiently has been proposed in \cite{tiurev_2022}. It is an open problem if this can be adapted to approximate the fidelity from the noisy stabilizer description. 

Finally, we point out that the methods and manipulations presented in this paper in a real network scenario require a classical protocol to synchronize and coordinate the network nodes and their actions. Depending on the nature of this classical layer, different classical waiting times arise. Those lead to different required memory times and hence asymmetric noise configurations. 

\appendices
\section{Bell pair via the X method from a 2D cluster}\label{a:x:protocol}
\noindent
Our goal here is to compute the required measurements using the $X$ method in a 2D grid taking a stairway path. We show that for any length of the stairway path, the required measurements are the corresponding $X$ measurements of the inner neighbors and six $Z$ measurements. 

We label the qubits in the path as $v_j$ with $j=1, \dots, l$, where $l-1$ is the length of the path, such that the target qubits are $v_1$ and $v_l$. Following the $X$ method, qubits $v_2,\dots, v_{l-1}$ have to be $X$-measured. Next, the resulting neighbors of $v_1$ and $v_l$ have to be $Z$-measured. Let us define $t$ as the $X$ measurement of qubit $v_{t+1}$ and $N_{v_j}^{(t)}$ is the neighborhood of $v_j$ after the $t$th $X$ measurement. In total, this method requires $(l-2)$ measurements in the $X$ basis and $|N_1^{(l-2)}\cup N_l^{(l-2)}|-2$ measurements in the $Z$ basis. 

From \cite{hahn_quantum_2019}, we use Eqs. (13)-(17) from the supplementary information. Moreover, we also make use that for a stairway path $|N_{v_j}^{(0)} \cap N_{v_{k}}^{(0)}|=2$ if $k=j+2, j-2$, and any other intersection between $N_{v_j}$ and $N_{v_k}$ is empty. Note that the non-empty intersections are unique. We look at the case for $l=5$ and using all of these equations we compute the final neighborhoods of $v_1$ and $v_5$. In Table~\ref{tab:x:method}, the initial neighborhoods are shown in the $t=0$ column, such that the outer neighbors of the path are labeled by $\{a, b, \dots, i\}$. Moreover, the neighborhoods of the qubits in the path at each step of the manipulation are shown.

\begin{table}[!t]
\centering
\caption{Neighborhoods of the qubits in a stairway path of length $5$ at the different steps of the $X$ method.}
\label{tab:x:method}
\begin{tabular}{l|l|l|l|l|}
\cline{2-5}
& $t=0$ & $t=1$ & $t=2$ & $t=3$ \\ \hline
\multicolumn{1}{|l|}{$N_{v_1}^{(t)}$} & $a, b, i, v_2$ & $v_3, g, h$ & $v_4, c, a, i$ & $v_5, h, f$ \\
\multicolumn{1}{|l|}{$N_{v_2}^{(t)}$} & $v_1, h, g, v_3$ & $\varnothing$ & $\varnothing$ & $\varnothing$ \\
\multicolumn{1}{|l|}{$N_{v_3}^{(t)}$} & $v_2, v_4, c, b$ & $v_1, v_4, c, a, i$ & $\varnothing$ & $\varnothing$ \\
\multicolumn{1}{|l|}{$N_{v_4}^{(t)}$} & $v_3, v_5, g, f$ & $v_3, v_5, g, f$ & $v_1, v_5, h, f$ & $\varnothing$ \\
\multicolumn{1}{|l|}{$N_{v_5}^{(t)}$} & $v_4, c, d, e$ & $v_4, c, d, e$ & $v_4, c, d, e$ & $v_1, d, e, a, i$ \\ \hline
\end{tabular}
\end{table}

From the last column in Table~\ref{tab:x:method}, we see that qubits $a, d, e, f, h, i$ have to be $Z$-measured to achieve a pair between $v_1$ and $v_5$. These qubits are the neighboring qubits of the path that only belong to one neighborhood of the qubits in the path in $t=0$. Note that the presented example has other cases for smaller $l$ included in it, e.g., the first step $t=1$ is an example for $l=3$. Thus, these results can be generalized for an arbitrary length of the stairway path, such that the following hold for any $l-2=n$ where $n\geq 0$ (can be proven via induction): $|N_{v_1}^{(n)}| = 4-g(n)$, $|N_{v_{n+2}}^{(n)}| = 4 + g(n)$, $|N_{v_1}^{(n)} \cap N_{v_{n+2}}^{(n)}| = 0$, $|N_{v_1}^{(n-1)} \cap N_{v_{n+2}}^{(n-1)}| = 2$. Thus, $|N_{v_1}^{(n)} \cup N_{v_{n+2}}^{(n)}| -2 = 6$, such that using the $X$ method in a stairway path only requires six $Z$ measurements.

\section{Three-qubit GHZ state as a target state}\label{a:3GHZ}
In this section, we describe the fidelity computation of a noisy $\text{GHZ}_3$ state and we give the generalized results for the same resource states as in Sec.~\ref{ssec:BP:NGHZ} - \ref{ssec:BP:tree}. In Table~\ref{tab:3GHZ} we give an overview of the fidelity in the LER for all the studied resource states.

\subsection{Fidelity computation}\label{ssec:fidelity:3GHZ}
Assume that the target state is between qubits $a$, $b$, and $c$, where $a$ and $c$ are the leaf qubits and $b$ is the root. Then, the set of noise maps that the measured qubits can take has 7 components, which are of the form
\begin{equation}\label{eq:3GHZ:maps}
    \mathcal{M}_{\alpha \beta \gamma}\rho' = p\rho' + \frac{1-p}{2}\left(\rho' + Z_a^{\alpha} Z_b^{\beta} Z_c^{\gamma} \rho' Z_a^{\alpha} Z_b^{\beta} Z_c^{\gamma}\right)
\end{equation}
where $\alpha, \beta, \gamma= 0,1$ and $\rho' = |{\text{GHZ}_3}\rangle\langle{\text{GHZ}_3}|$ is the noiseless target state, and $\mathcal{M}_{000}$ is not in the set. We define a new weight vector,
$\boldsymbol{w}=(w_{001}, w_{010}, w_{011}, w_{100}, w_{101}, w_{110}, w_{111})$. Using the fact that any $\mathcal{M}_{\alpha\beta\gamma}$ applied $w_{\alpha\beta\gamma}$ times with probability $p$ is the same as applying it once with probability $p^{w_{\alpha\beta\gamma}}$, the expression for the fidelity of a $\text{GHZ}_3$ state is
\begin{equation}\label{eq:general:fidelity:3GHZ}
\begin{aligned}
    F(p,\boldsymbol{w})= \frac{1}{4}(1 &+ p^{2 + w_{001} + w_{011} + w_{100} + w_{110}} \\ &+ p^{3 + w_{001} + w_{010} + w_{100} + w_{111}} \\ &+ p^{3 + w_{010} + w_{011} + w_{110} + w_{111}}).
\end{aligned}
\end{equation}
This is valid for any resource state that has been manipulated via local Pauli measurements into a $\text{GHZ}_3$ state. The general case including full-merging operations can be computed following the Bell pair case. The fidelity in the LER, where $1-p=\epsilon$ for small $\epsilon$ is 
\begin{equation}\label{eq:approx:fidelity:3GHZ}
    F(\epsilon, \boldsymbol{w})\simeq 1-\frac{1}{2}\left(4+ \sum_{\alpha,\beta,\gamma=0}^1w_{\alpha\beta\gamma}-w_{000}-w_{101}\right)\epsilon.
\end{equation}
Given that the sum of the elements of $\boldsymbol{w}$ corresponds to the number of performed local Pauli measurements, Eq.~\eqref{eq:approx:fidelity:3GHZ} can be computed with a counting argument as long as $w_{101}=0$, which corresponds to the case where the leaf qubits do not share neighbors, and are thus, independent of each other.

\subsection{Resource: N-qubit GHZ state}\label{ssec:3GHZ:NGHZ}
Assume that the root of the target state is the same as the root of the resource state. Thus, only the non-target leaves have to be $Z$-measured, such that $\boldsymbol{w}=(0,N-3,0,0,0,0,0)$.

\subsection{Resource: 1D cluster}\label{ssec:3GHZ:1D}
We connect the root qubit with each of the leaves using the shortest path, such that the two paths have $n_{1\text{D}}^{(ab)}$ and $n_{1\text{D}}^{(bc)}$ inner neighbors each, and the total is $n_{1\text{D}}^{(abc)} = n_{1\text{D}}^{(ab)}+n_{1\text{D}}^{(bc)}$. We use the $Y$ method and consider that the two leaf qubits do not share outer neighbors such that their noise is not correlated, leading to $w_{101}=w_{111}=0$. We consider an outwards side-to-side strategy, such that the qubits in the two paths are measured in the direction towards the leaf qubit. Note that the weight vector for any other strategy can be easily computed. The weight vector is
\begin{equation}\label{eq:1D:3GHZ:outward}
\begin{aligned}
    w_{001} &= 1+\frac{1}{2}\left[n_{1\text{D}}^{(bc)}-g\left(n_{1\text{D}}^{(bc)}\right)\right], \\ 
    w_{010}  &= 0, \\
    w_{011}  &= \frac{1}{2}\left[n_{1\text{D}}^{(bc)}+g\left(n_{1\text{D}}^{(bc)}\right)\right],  \\ 
    w_{100}  &= 1+\frac{1}{2}\left[n_{1\text{D}}^{(ab)}-g\left(n_{1\text{D}}^{(ab)}\right)\right], \\
    w_{110}  &= \frac{1}{2}\left[n_{1\text{D}}^{(ab)}+g\left(n_{1\text{D}}^{(ab)}\right)\right].
\end{aligned}
\end{equation}

\subsection{Resource: 2D cluster}\label{ssec:3GHZ:2D}
We use the same assumptions as for the 1D case with now two paths of $n_{2\text{D}}^{(ab)}$ and $n_{2\text{D}}^{(bc)}$ inner neighbors correspondingly, and a total of $n_{2\text{D}}^{(abc)} = n_{2\text{D}}^{(ab)}+n_{2\text{D}}^{(bc)}$. Additionally, we consider the lower-bound approximation that the shortest paths are straight lines, as in Sec.~\ref{ssec:BP:2D}. The corresponding weight vector is $\boldsymbol{w}_{2\text{D}} = 3\boldsymbol{w}_{1\text{D}} + (0,2,0,0,0,0,0)$, where $\boldsymbol{w}_{1\text{D}}$ corresponds to the weight vector for the 1D cluster (for any order of measurements) using $n_{2\text{D}}^{(ab)}$ and $n_{2\text{D}}^{(bc)}$.

\subsection{Resource: kD cluster}\label{ssec:3GHZ:kD}
We use the same assumptions as for the 2D case with now two paths of $n_{k\text{D}}^{(ab)}$ and $n_{k\text{D}}^{(bc)}$ inner neighbors correspondingly, and a total of $n_{k\text{D}}^{(abc)} = n_{k\text{D}}^{(ab)}+n_{k\text{D}}^{(bc)}$. The corresponding weight vector is $\boldsymbol{w}_{k\text{D}}=(2k-1)\boldsymbol{w}_{1\text{D}}+(0,2k-2,0,0,0,0,0)$, where $\boldsymbol{w}_{1\text{D}}$ corresponds to the weight vector for the 1D cluster (for any order of measurements) using $n_{k\text{D}}^{(ab)}$ and $n_{k\text{D}}^{(bc)}$. 

\subsection{Resource: Binary tree cluster}
The three target qubits are in arbitrary positions in the binary tree cluster. Before reaching a $\text{GHZ}_3$ state, we need to reach a four-qubit GHZ state, where all the target qubits are leaves. This first step can be achieved by following the $Y$ method for the 3 paths that are there. Each of these paths goes from one of the target qubits ($a$, $b$, and $c$) to the root qubit of this intermediate state, which we label as $r$. These paths have $n_{ar}$, $n_{br}$ and $n_{cr}$ inner neighbors each, such that the total number of inner neighbors is $n_{\text{tree}}^{(abc)}=n_{ar} +n_{br}+n_{cr}$. Now, to reach the final $\text{GHZ}_3$ state, qubit $r$ is $X$-measured, where $b$ is chosen as the special neighbor to achieve the structure where $b$ is the root. Considering that the $Y$ method in the first step has been performed following the outward side-to-side strategy, the resulting weight vector is 
\begin{equation}
    \begin{aligned}
    w_{001} &= n_{ar}-g(n_{ar})+2, &w_{011} &= n_{cr}+g(n_{cr}), \\ w_{010} &= n_{br}-g(n_{br})+3, &w_{110} &= n_{ar}+g(n_{ar}), \\ 
    w_{100} &= n_{cr}-g(n_{cr})+2, &w_{110} &= n_{br}+g(n_{br}), \\
    w_{101} &= 0.
    \end{aligned}
\end{equation}

\begin{table}[!t]
\caption{Fidelity in the LER for a $\text{GHZ}_3$ state between any three nodes in a network of $N$ nodes depending on the resource state.}
\label{tab:3GHZ}
\centering
\begin{tabular}{|l|l|}
\hline
\textbf{Resource state} & \textbf{Fidelity in the LER} \\ \hline \hline
$\text{GHZ}_N$ state & $1-\frac{1}{2}(N-1)\epsilon$ \\ 
1D cluster & $1-\frac{1}{2}\left(6 + n_{1\text{D}}^{(abc)}\right)\epsilon$ \\ 
2D cluster & $1-\frac{1}{2}\left(12 + 3n_{2\text{D}}^{(abc)}\right)\epsilon$ \\ 
$k$D cluster & $1-\frac{1}{2}\left[(2k-1)(n_{k\text{D}}^{(abc)} + 3) +3\right]\epsilon$ \\ 
Binary tree cluster & $1-\frac{1}{2}\left(11+2n_{\text{tree}}^{(abc)}\right)\epsilon$ \\ \hline
\end{tabular}
\end{table}

\bibliographystyle{IEEEtran}
\bibliography{refs_JSAC_v2.bib}

\begin{thebibliography}{10}
\providecommand{\url}[1]{#1}
\csname url@samestyle\endcsname
\providecommand{\newblock}{\relax}
\providecommand{\bibinfo}[2]{#2}
\providecommand{\BIBentrySTDinterwordspacing}{\spaceskip=0pt\relax}
\providecommand{\BIBentryALTinterwordstretchfactor}{4}
\providecommand{\BIBentryALTinterwordspacing}{\spaceskip=\fontdimen2\font plus
\BIBentryALTinterwordstretchfactor\fontdimen3\font minus
  \fontdimen4\font\relax}
\providecommand{\BIBforeignlanguage}[2]{{%
\expandafter\ifx\csname l@#1\endcsname\relax
\typeout{** WARNING: IEEEtran.bst: No hyphenation pattern has been}%
\typeout{** loaded for the language `#1'. Using the pattern for}%
\typeout{** the default language instead.}%
\else
\language=\csname l@#1\endcsname
\fi
#2}}
\providecommand{\BIBdecl}{\relax}
\BIBdecl

\bibitem{Kimble2008}
H.~J. Kimble, ``The quantum internet,'' \emph{Nature}, vol. 453, no. 7198, pp.
  1023--1030, 2008.

\bibitem{wehner_internet}
S.~Wehner, D.~Elkouss, and R.~Hanson, ``Quantum internet: A vision for the road
  ahead,'' \emph{Science}, vol. 362, no. 6412, p. eaam9288, 2018.

\bibitem{Azuma2021}
K.~Azuma, S.~B{\"a}uml, T.~Coopmans, D.~Elkouss, and B.~Li, ``Tools for quantum
  network design,'' \emph{AVS Quantum Science}, vol.~3, no.~1, p. 014101, 2021.

\bibitem{azuma_2022}
K.~Azuma, S.~E. Economou, D.~Elkouss, P.~Hilaire, L.~Jiang, H.-K. Lo, and
  I.~Tzitrin, ``Quantum repeaters: From quantum networks to the quantum
  internet,'' \emph{Rev. Mod. Phys.}, vol.~95, p. 045006, Dec 2023.

\bibitem{Acin2018Quantum}
A.~Ac{\'\i}n, I.~Bloch, H.~Buhrman, T.~Calarco, C.~Eichler, J.~Eisert,
  D.~Esteve, N.~Gisin, S.~J. Glaser, F.~Jelezko \emph{et~al.}, ``The quantum
  technologies roadmap: a european community view,'' \emph{New J. Phys.},
  vol.~20, no.~8, p. 080201, 2018.

\bibitem{Riedel2019Europe}
M.~Riedel, M.~Kovacs, P.~Zoller, J.~Mlynek, and T.~Calarco, ``Europe’s
  quantum flagship initiative,'' \emph{Quantum Sci. Technol.}, vol.~4, no.~2,
  p. 020501, 2019.

\bibitem{Cacciapuoti2020}
A.~S. Cacciapuoti, M.~Caleffi, F.~Tafuri, F.~S. Cataliotti, S.~Gherardini, and
  G.~Bianchi, ``Quantum internet: Networking challenges in distributed quantum
  computing,'' \emph{IEEE Network}, vol.~34, no.~1, pp. 137--143, 2020.

\bibitem{Eldredge2018}
Z.~Eldredge, M.~Foss-Feig, J.~A. Gross, S.~L. Rolston, and A.~V. Gorshkov,
  ``Optimal and secure measurement protocols for quantum sensor networks,''
  \emph{Phys. Rev. A}, vol.~97, p. 042337, Apr 2018.

\bibitem{Sekatski2020}
P.~Sekatski, S.~W\"olk, and W.~D\"ur, ``Optimal distributed sensing in noisy
  environments,'' \emph{Phys. Rev. Res.}, vol.~2, p. 023052, Apr 2020.

\bibitem{CiracDistributed}
J.~I. Cirac, A.~K. Ekert, S.~F. Huelga, and C.~Macchiavello, ``Distributed
  quantum computation over noisy channels,'' \emph{Phys. Rev. A}, vol.~59, pp.
  4249--4254, Jun 1999.

\bibitem{caleffi_2018}
M.~Caleffi, A.~S. Cacciapuoti, and G.~Bianchi, ``Quantum internet: From
  communication to distributed computing!'' in \emph{Proceedings of the 5th ACM
  International Conference on Nanoscale Computing and Communication}, ser.
  NANOCOM '18.\hskip 1em plus 0.5em minus 0.4em\relax New York, NY, USA:
  Association for Computing Machinery, 2018.

\bibitem{Gisin2002}
N.~Gisin, G.~Ribordy, W.~Tittel, and H.~Zbinden, ``Quantum cryptography,''
  \emph{Rev. Mod. Phys.}, vol.~74, pp. 145--195, Mar 2002.

\bibitem{ShorSimple}
P.~W. Shor and J.~Preskill, ``Simple proof of security of the bb84 quantum key
  distribution protocol,'' \emph{Phys. Rev. Lett.}, vol.~85, pp. 441--444, Jul
  2000.

\bibitem{Murta2020Quantum}
G.~Murta, F.~Grasselli, H.~Kampermann, and D.~Bru{\ss}, ``Quantum conference
  key agreement: A review,'' \emph{Adv. Quantum Technol.}, vol.~3, no.~11, p.
  2000025, 2020.

\bibitem{hahn_anonymous}
F.~Hahn, J.~de~Jong, and A.~Pappa, ``Anonymous quantum conference key
  agreement,'' \emph{PRX Quantum}, vol.~1, p. 020325, Dec 2020.

\bibitem{markham_graph}
D.~Markham and B.~C. Sanders, ``Graph states for quantum secret sharing,''
  \emph{Phys. Rev. A}, vol.~78, p. 042309, Oct 2008.

\bibitem{Hillery1999}
M.~Hillery, V.~Bu\ifmmode~\check{z}\else \v{z}\fi{}ek, and A.~Berthiaume,
  ``Quantum secret sharing,'' \emph{Phys. Rev. A}, vol.~59, pp. 1829--1834, Mar
  1999.

\bibitem{Hillery2006}
M.~Hillery, M.~Ziman, V.~Bužek, and M.~Bieliková, ``Towards quantum-based
  privacy and voting,'' \emph{Phys. Lett. A}, vol. 349, no.~1, pp. 75--81,
  2006.

\bibitem{Vaccaro2007}
J.~A. Vaccaro, J.~Spring, and A.~Chefles, ``Quantum protocols for anonymous
  voting and surveying,'' \emph{Phys. Rev. A}, vol.~75, p. 012333, Jan 2007.

\bibitem{BennetTeleporting}
C.~H. Bennett, G.~Brassard, C.~Cr\'epeau, R.~Jozsa, A.~Peres, and W.~K.
  Wootters, ``Teleporting an unknown quantum state via dual classical and
  einstein-podolsky-rosen channels,'' \emph{Phys. Rev. Lett.}, vol.~70, pp.
  1895--1899, Mar 1993.

\bibitem{Briegel1998}
H.-J. Briegel, W.~D{\"u}r, J.~I. Cirac, and P.~Zoller, ``Quantum repeaters: The
  role of imperfect local operations in quantum communication,'' \emph{Phys.
  Rev. Lett.}, vol.~81, pp. 5932--5935, Dec 1998.

\bibitem{Sangouard2011}
N.~Sangouard, C.~Simon, H.~de~Riedmatten, and N.~Gisin, ``Quantum repeaters
  based on atomic ensembles and linear optics,'' \emph{Rev. Mod. Phys.},
  vol.~83, pp. 33--80, Mar 2011.

\bibitem{Munro2015}
W.~J. Munro, K.~Azuma, K.~Tamaki, and K.~Nemoto, ``Inside quantum repeaters,''
  \emph{IEEE J. Sel. Top. Quantum Electron.}, vol.~21, no.~3, pp. 78--90, 2015.

\bibitem{Krutyanskiy2023}
V.~Krutyanskiy, M.~Canteri, M.~Meraner, J.~Bate, V.~Krcmarsky, J.~Schupp,
  N.~Sangouard, and B.~P. Lanyon, ``Telecom-wavelength quantum repeater node
  based on a trapped-ion processor,'' \emph{Phys. Rev. Lett.}, vol. 130, p.
  213601, May 2023.

\bibitem{Pompili2021}
M.~Pompili, S.~L.~N. Hermans, S.~Baier, H.~K.~C. Beukers, P.~C. Humphreys,
  R.~N. Schouten, R.~F.~L. Vermeulen, M.~J. Tiggelman, L.~dos Santos~Martins,
  B.~Dirkse, S.~Wehner, and R.~Hanson, ``Realization of a multinode quantum
  network of remote solid-state qubits,'' \emph{Science}, vol. 372, no. 6539,
  pp. 259--264, 2021.

\bibitem{chen2021integrated}
Y.-A. Chen, Q.~Zhang, T.-Y. Chen, W.-Q. Cai, S.-K. Liao, J.~Zhang, K.~Chen,
  J.~Yin, J.-G. Ren, Z.~Chen \emph{et~al.}, ``An integrated space-to-ground
  quantum communication network over 4,600 kilometres,'' \emph{Nature}, vol.
  589, no. 7841, pp. 214--219, 2021.

\bibitem{li2019experimental}
Z.-D. Li, R.~Zhang, X.-F. Yin, L.-Z. Liu, Y.~Hu, Y.-Q. Fang, Y.-Y. Fei,
  X.~Jiang, J.~Zhang, L.~Li \emph{et~al.}, ``Experimental quantum repeater
  without quantum memory,'' \emph{Nat. Photonics}, vol.~13, no.~9, pp.
  644--648, 2019.

\bibitem{hermans2022qubit}
S.~Hermans, M.~Pompili, H.~Beukers, S.~Baier, J.~Borregaard, and R.~Hanson,
  ``Qubit teleportation between non-neighbouring nodes in a quantum network,''
  \emph{Nature}, vol. 605, no. 7911, pp. 663--668, 2022.

\bibitem{langenfeld2021quantum}
S.~Langenfeld, S.~Welte, L.~Hartung, S.~Daiss, P.~Thomas, O.~Morin,
  E.~Distante, and G.~Rempe, ``Quantum teleportation between remote qubit
  memories with only a single photon as a resource,'' \emph{Phys. Rev. Lett.},
  vol. 126, no.~13, p. 130502, 2021.

\bibitem{CacciapuotiWhen}
A.~S. Cacciapuoti, M.~Caleffi, R.~Van~Meter, and L.~Hanzo, ``When entanglement
  meets classical communications: Quantum teleportation for the quantum
  internet,'' \emph{IEEE Trans. Commun.}, vol.~68, no.~6, pp. 3808--3833, 2020.

\bibitem{pirker_quantum_2019}
A.~Pirker and W.~Dür, ``A quantum network stack and protocols for reliable
  entanglement-based networks,'' \emph{New J. Phys.}, vol.~21, no.~3, p.
  033003, Mar. 2019.

\bibitem{vanmeter_recursive}
R.~V. Meter, J.~Touch, and C.~Horsman, ``\BIBforeignlanguage{en}{Recursive
  quantum repeater networks},'' \emph{\BIBforeignlanguage{en}{Prog. Inform.}},
  no.~8, p.~65, mar 2011.

\bibitem{Muralidharan2016Optimal}
S.~Muralidharan, L.~Li, J.~Kim, N.~L{\"u}tkenhaus, M.~D. Lukin, and L.~Jiang,
  ``Optimal architectures for long distance quantum communication,'' \emph{Sci.
  Rep.}, vol.~6, no.~1, p. 20463, 2016.

\bibitem{bugalho2023distributing}
L.~Bugalho, B.~C. Coutinho, F.~A. Monteiro, and Y.~Omar, ``Distributing
  multipartite entanglement over noisy quantum networks,'' \emph{Quantum},
  vol.~7, p. 920, 2023.

\bibitem{patil2022entanglement}
A.~Patil, M.~Pant, D.~Englund, D.~Towsley, and S.~Guha, ``Entanglement
  generation in a quantum network at distance-independent rate,'' \emph{Npj
  Quantum Inf.}, vol.~8, no.~1, p.~51, 2022.

\bibitem{fischer_2021}
A.~Fischer and D.~Towsley, ``Distributing graph states across quantum
  networks,'' in \emph{2021 IEEE International Conference on Quantum Computing
  and Engineering (QCE)}, 2021, pp. 324--333.

\bibitem{pant2019routing}
M.~Pant, H.~Krovi, D.~Towsley, L.~Tassiulas, L.~Jiang, P.~Basu, D.~Englund, and
  S.~Guha, ``Routing entanglement in the quantum internet,'' \emph{Npj Quantum
  Inf.}, vol.~5, no.~1, p.~25, 2019.

\bibitem{matsuzaki_2010}
Y.~Matsuzaki, S.~C. Benjamin, and J.~Fitzsimons, ``Probabilistic growth of
  large entangled states with low error accumulation,'' \emph{Phys. Rev.
  Lett.}, vol. 104, p. 050501, Feb 2010.

\bibitem{epping2016large}
M.~Epping, H.~Kampermann, and D.~Bru{\ss}, ``Large-scale quantum networks based
  on graphs,'' \emph{New J. Phys.}, vol.~18, no.~5, p. 053036, 2016.

\bibitem{pirandola2019end}
S.~Pirandola, ``End-to-end capacities of a quantum communication network,''
  \emph{Commun. Phys.}, vol.~2, no.~1, p.~51, 2019.

\bibitem{wallnofer_simulating_2022}
J.~Walln{\"o}fer, F.~Hahn, M.~G{\"u}ndo{\u{g}}an, J.~S. Sidhu, F.~Wiesner,
  N.~Walk, J.~Eisert, and J.~Wolters, ``Simulating quantum repeater strategies
  for multiple satellites,'' \emph{Commun. Phys.}, vol.~5, no.~1, p. 169, Jun.
  2022.

\bibitem{kozlowski_2019}
W.~Kozlowski and S.~Wehner, ``Towards large-scale quantum networks,'' in
  \emph{Proceedings of the Sixth Annual ACM International Conference on
  Nanoscale Computing and Communication}, ser. NANOCOM '19.\hskip 1em plus
  0.5em minus 0.4em\relax New York, NY, USA: Association for Computing
  Machinery, 2019.

\bibitem{coutinho2022robustness}
B.~Coelho~Coutinho, W.~J. Munro, K.~Nemoto, and Y.~Omar, ``Robustness of noisy
  quantum networks,'' \emph{Commun. Phys.}, vol.~5, no.~1, p. 105, 2022.

\bibitem{Azuma2015All}
K.~Azuma, K.~Tamaki, and H.-K. Lo, ``All-photonic quantum repeaters,''
  \emph{Nat. Commun.}, vol.~6, no.~1, p. 6787, 2015.

\bibitem{coopmans_netsquid_2021}
T.~Coopmans, R.~Knegjens, A.~Dahlberg, D.~Maier, L.~Nijsten, J.~d.~O. Filho,
  M.~Papendrecht, J.~Rabbie, F.~Rozp\k{e}dek, M.~Skrzypczyk, L.~Wubben,
  W.~de~Jong, D.~Podareanu, A.~Torres-Knoop, D.~Elkouss, and S.~Wehner,
  ``{NetSquid}, a {NETwork} {Simulator} for {QUantum} {Information} using
  {Discrete} events,'' \emph{Commun. Phys.}, vol.~4, no.~1, p. 164, Dec. 2021.

\bibitem{Pirandola2017Fundamental}
S.~Pirandola, R.~Laurenza, C.~Ottaviani, and L.~Banchi, ``Fundamental limits of
  repeaterless quantum communications,'' \emph{Nat. Commun.}, vol.~8, no.~1, p.
  15043, 2017.

\bibitem{Guha2015}
S.~Guha, H.~Krovi, C.~A. Fuchs, Z.~Dutton, J.~A. Slater, C.~Simon, and
  W.~Tittel, ``Rate-loss analysis of an efficient quantum repeater
  architecture,'' \emph{Phys. Rev. A}, vol.~92, p. 022357, Aug 2015.

\bibitem{Sanogouard2009}
N.~Sangouard, R.~Dubessy, and C.~Simon, ``Quantum repeaters based on single
  trapped ions,'' \emph{Phys. Rev. A}, vol.~79, p. 042340, Apr 2009.

\bibitem{pirker_modular_2018}
A.~Pirker, J.~Wallnöfer, and W.~Dür, ``\BIBforeignlanguage{en}{Modular
  architectures for quantum networks},'' \emph{\BIBforeignlanguage{en}{New J.
  Phys.}}, vol.~20, no.~5, p. 053054, May 2018.

\bibitem{meignant_2019}
C.~Meignant, D.~Markham, and F.~Grosshans, ``Distributing graph states over
  arbitrary quantum networks,'' \emph{Phys. Rev. A}, vol. 100, p. 052333, Nov
  2019.

\bibitem{gyongyosi2019opportunistic}
L.~Gyongyosi and S.~Imre, ``Opportunistic entanglement distribution for the
  quantum internet,'' \emph{Sci. Rep.}, vol.~9, no.~1, p. 2219, 2019.

\bibitem{miguel-ramiro_optimized_2021}
J.~Miguel-Ramiro, A.~Pirker, and W.~D{\"{u}}r, ``Optimized {Q}uantum
  {N}etworks,'' \emph{{Quantum}}, vol.~7, p. 919, Feb. 2023.

\bibitem{dur_stability_2004}
W.~D{\"u}r and H.-J. Briegel, ``Stability of macroscopic entanglement under
  decoherence,'' \emph{Phys. Rev. Lett.}, vol.~92, p. 180403, May 2004.

\bibitem{hein_entanglement_2005}
M.~Hein, W.~D{\"u}r, and H.-J. Briegel, ``Entanglement properties of
  multipartite entangled states under the influence of decoherence,''
  \emph{Phys. Rev. A}, vol.~71, p. 032350, Mar 2005.

\bibitem{caleffi_2020}
M.~Caleffi and A.~S. Cacciapuoti, ``Quantum switch for the quantum internet:
  Noiseless communications through noisy channels,'' \emph{IEEE J. Sel. Areas
  Commun.}, vol.~38, no.~3, pp. 575--588, 2020.

\bibitem{chiribella2021indefinite}
G.~Chiribella, M.~Banik, S.~S. Bhattacharya, T.~Guha, M.~Alimuddin, A.~Roy,
  S.~Saha, S.~Agrawal, and G.~Kar, ``Indefinite causal order enables perfect
  quantum communication with zero capacity channels,'' \emph{New J. Phys.},
  vol.~23, no.~3, p. 033039, 2021.

\bibitem{caleffi2023beyond}
M.~Caleffi, K.~Simonov, and A.~S. Cacciapuoti, ``Beyond shannon limits: Quantum
  communications through quantum paths,'' \emph{IEEE J. Sel. Areas Commun.},
  2023.

\bibitem{Cuquet2012}
M.~Cuquet and J.~Calsamiglia, ``Growth of graph states in quantum networks,''
  \emph{Phys. Rev. A}, vol.~86, p. 042304, Oct 2012.

\bibitem{vardoyan_stochastic_2019}
G.~Vardoyan, S.~Guha, P.~Nain, and D.~Towsley, ``On the stochastic analysis of
  a quantum entanglement distribution switch,'' \emph{IEEE Trans. Quantum
  Engineering}, vol.~2, pp. 1--16, 2021.

\bibitem{Vardoyan2020}
------, ``On the exact analysis of an idealized quantum switch,''
  \emph{Perform. Evaluation}, vol. 144, p. 102141, 2020.

\bibitem{nain_2020}
P.~Nain, G.~Vardoyan, S.~Guha, and D.~Towsley, ``On the analysis of a
  multipartite entanglement distribution switch,'' \emph{Proc. ACM Meas. Anal.
  Comput. Syst.}, vol.~4, no.~2, jun 2020.

\bibitem{nain2022analysis}
------, ``Analysis of a tripartite entanglement distribution switch,''
  \emph{Queueing Syst.}, vol. 101, no. 3-4, pp. 291--328, 2022.

\bibitem{epping_robust_2016}
M.~Epping, H.~Kampermann, and D.~Bruß, ``\BIBforeignlanguage{en}{Robust
  entanglement distribution via quantum network coding},''
  \emph{\BIBforeignlanguage{en}{New J. Phys.}}, vol.~18, no.~10, p. 103052,
  Oct. 2016.

\bibitem{leung_butterfly}
D.~Leung, J.~Oppenheim, and A.~Winter, ``Quantum network communication—the
  butterfly and beyond,'' \emph{IEEE Trans. Inf. Theory}, vol.~56, no.~7, pp.
  3478--3490, 2010.

\bibitem{hahn_quantum_2019}
F.~Hahn, A.~Pappa, and J.~Eisert, ``\BIBforeignlanguage{en}{Quantum network
  routing and local complementation},'' \emph{\BIBforeignlanguage{en}{Npj
  Quantum Inf.}}, vol.~5, no.~1, pp. 1--7, Sep. 2019.

\bibitem{Satoh2016}
T.~Satoh, K.~Ishizaki, S.~Nagayama, and R.~Van~Meter, ``Analysis of quantum
  network coding for realistic repeater networks,'' \emph{Phys. Rev. A},
  vol.~93, p. 032302, Mar 2016.

\bibitem{jong_2022}
J.~de~Jong, F.~Hahn, N.~Tcholtchev, M.~Hauswirth, and A.~Pappa, ``Extracting
  maximal entanglement from linear cluster states,'' \emph{arXiv:2211.16758
  [quant-ph]}, 2022.

\bibitem{Frowis2018}
F.~Fr\"owis, P.~Sekatski, W.~D{\"u}r, N.~Gisin, and N.~Sangouard, ``Macroscopic
  quantum states: Measures, fragility, and implementations,'' \emph{Rev. Mod.
  Phys.}, vol.~90, p. 025004, May 2018.

\bibitem{hein_multiparty_2004}
M.~Hein, J.~Eisert, and H.~J. Briegel, ``Multiparty entanglement in graph
  states,'' \emph{Phys. Rev. A}, vol.~69, no.~6, p. 062311, Jun. 2004.

\bibitem{hein_entanglement_2006}
M.~Hein, W.~D{\"u}r, J.~Eisert, R.~Raussendorf, M.~Van~den Nest, and H.-J.
  Briegel, ``Entanglement in graph states and its applications,'' in
  \emph{Quantum Computers, Algorithms and Chaos}, ser. Proceedings of the
  International School of Physics "Enrico Fermi", G.~Casati, D.~L.
  Shepelyansky, P.~Zoller, and G.~Benenti, Eds., vol. 162.\hskip 1em plus 0.5em
  minus 0.4em\relax IOS Press, 2006, p. 115–218.

\bibitem{mor_noisy}
M.~F. Mor-Ruiz and W.~D{\"u}r, ``Noisy stabilizer formalism,'' \emph{Phys. Rev.
  A}, vol. 107, p. 032424, Mar 2023.

\bibitem{gyongyosi2017entanglement}
L.~Gyongyosi and S.~Imre, ``Entanglement-gradient routing for quantum
  networks,'' \emph{Sci. Rep.}, vol.~7, no.~1, p. 14255, 2017.

\bibitem{miguel2020delocalized}
J.~Miguel-Ramiro and W.~D{\"u}r, ``Delocalized information in quantum
  networks,'' \emph{New J. Phys.}, vol.~22, no.~4, p. 043011, 2020.

\bibitem{schoute2016shortcuts}
E.~Schoute, L.~Mancinska, T.~Islam, I.~Kerenidis, and S.~Wehner, ``Shortcuts to
  quantum network routing,'' 2016, arXiv preprint arXiv:1610.05238.

\bibitem{osi_1980}
H.~Zimmermann, ``Osi reference model - the iso model of architecture for open
  systems interconnection,'' \emph{IEEE Trans. Commun.}, vol.~28, no.~4, pp.
  425--432, 1980.

\bibitem{mannalath}
V.~Mannalath and A.~Pathak, ``Multiparty entanglement routing in quantum
  networks,'' 2022.

\bibitem{hahn_limitations}
F.~Hahn, A.~Dahlberg, J.~Eisert, and A.~Pappa, ``Limitations of
  nearest-neighbor quantum networks,'' \emph{Phys. Rev. A}, vol. 106, p.
  L010401, Jul 2022.

\bibitem{raussendorf_measurement}
R.~Raussendorf, D.~E. Browne, and H.~J. Briegel, ``Measurement-based quantum
  computation on cluster states,'' \emph{Phys. Rev. A}, vol.~68, p. 022312, Aug
  2003.

\bibitem{raussendorf_one}
R.~Raussendorf and H.~J. Briegel, ``A one-way quantum computer,'' \emph{Phys.
  Rev. Lett.}, vol.~86, pp. 5188--5191, May 2001.

\bibitem{dahlberg_transforming}
A.~Dahlberg, J.~Helsen, and S.~Wehner, ``Transforming graph states to
  {B}ell-pairs is {NP}-{C}omplete,'' \emph{{Quantum}}, vol.~4, p. 348, oct
  2020.

\bibitem{GottesmanThesis}
D.~Gottesman, ``Stabilizer codes and quantum error correction,'' Ph.D.
  dissertation, California Institute of Technology, Pasadena, California, 1997.

\bibitem{nest_invariants_2005}
M.~Van~den Nest, J.~Dehaene, and B.~De~Moor, ``The invariants of the local
  {Clifford} group,'' \emph{Phys. Rev . A}, vol.~71, no.~2, p. 022310, Feb.
  2005.

\bibitem{matsuo2018}
T.~Matsuo, T.~Satoh, S.~Nagayama, and R.~Van~Meter, ``Analysis of
  measurement-based quantum network coding over repeater networks under noisy
  conditions,'' \emph{Phys. Rev. A}, vol.~97, p. 062328, Jun 2018.

\bibitem{wallnoefer_meas}
J.~Walln{\"o}fer and W.~D{\"u}r, ``Measurement-based quantum communication with
  resource states generated by entanglement purification,'' \emph{Phys. Rev.
  A}, vol.~95, p. 012303, Jan 2017.

\bibitem{dur2005standard}
W.~D\"ur, M.~Hein, J.~I. Cirac, and H.-J. Briegel, ``Standard forms of noisy
  quantum operations via depolarization,'' \emph{Phys. Rev. A}, vol.~72, p.
  052326, Nov 2005.

\bibitem{ekert_1991}
A.~K. Ekert, ``Quantum cryptography based on bell's theorem,'' \emph{Phys. Rev.
  Lett.}, vol.~67, pp. 661--663, Aug 1991.

\bibitem{benor_2005}
M.~Ben-Or and A.~Hassidim, ``Fast quantum byzantine agreement,'' in
  \emph{Proceedings of the Thirty-Seventh Annual ACM Symposium on Theory of
  Computing}, ser. STOC '05.\hskip 1em plus 0.5em minus 0.4em\relax New York,
  NY, USA: Association for Computing Machinery, 2005, p. 481–485.

\bibitem{shettell_graph}
N.~Shettell and D.~Markham, ``Graph states as a resource for quantum
  metrology,'' \emph{Phys. Rev. Lett.}, vol. 124, p. 110502, Mar 2020.

\bibitem{RevModPhys.90.035005}
L.~Pezz{\`e}, A.~Smerzi, M.~K. Oberthaler, R.~Schmied, and P.~Treutlein,
  ``Quantum metrology with nonclassical states of atomic ensembles,''
  \emph{Rev. Mod. Phys.}, vol.~90, p. 035005, Sep 2018.

\bibitem{giovannetti2011advances}
V.~Giovannetti, S.~Lloyd, and L.~Maccone, ``Advances in quantum metrology,''
  \emph{Nat. Photonics}, vol.~5, no.~4, p. 222, 2011.

\bibitem{T_th_2014}
G.~T{\'{o}}th and I.~Apellaniz, ``Quantum metrology from a quantum information
  science perspective,'' \emph{J. Phys. A}, vol.~47, no.~42, p. 424006, oct
  2014.

\bibitem{bravyi}
S.~Bravyi, Y.~Sharma, M.~Szegedy, and R.~de~Wolf, ``Generating $k$ epr-pairs
  from an n-party resource state,'' 2022.

\bibitem{tiurev_2022}
K.~Tiurev and A.~S. S\o{}rensen, ``Fidelity measurement of a multiqubit cluster
  state with minimal effort,'' \emph{Phys. Rev. Res.}, vol.~4, p. 033162, Aug
  2022.

\end{thebibliography}

\newpage

\begin{IEEEbiography}[{\includegraphics[width=1in,height=1.25in,clip,keepaspectratio]{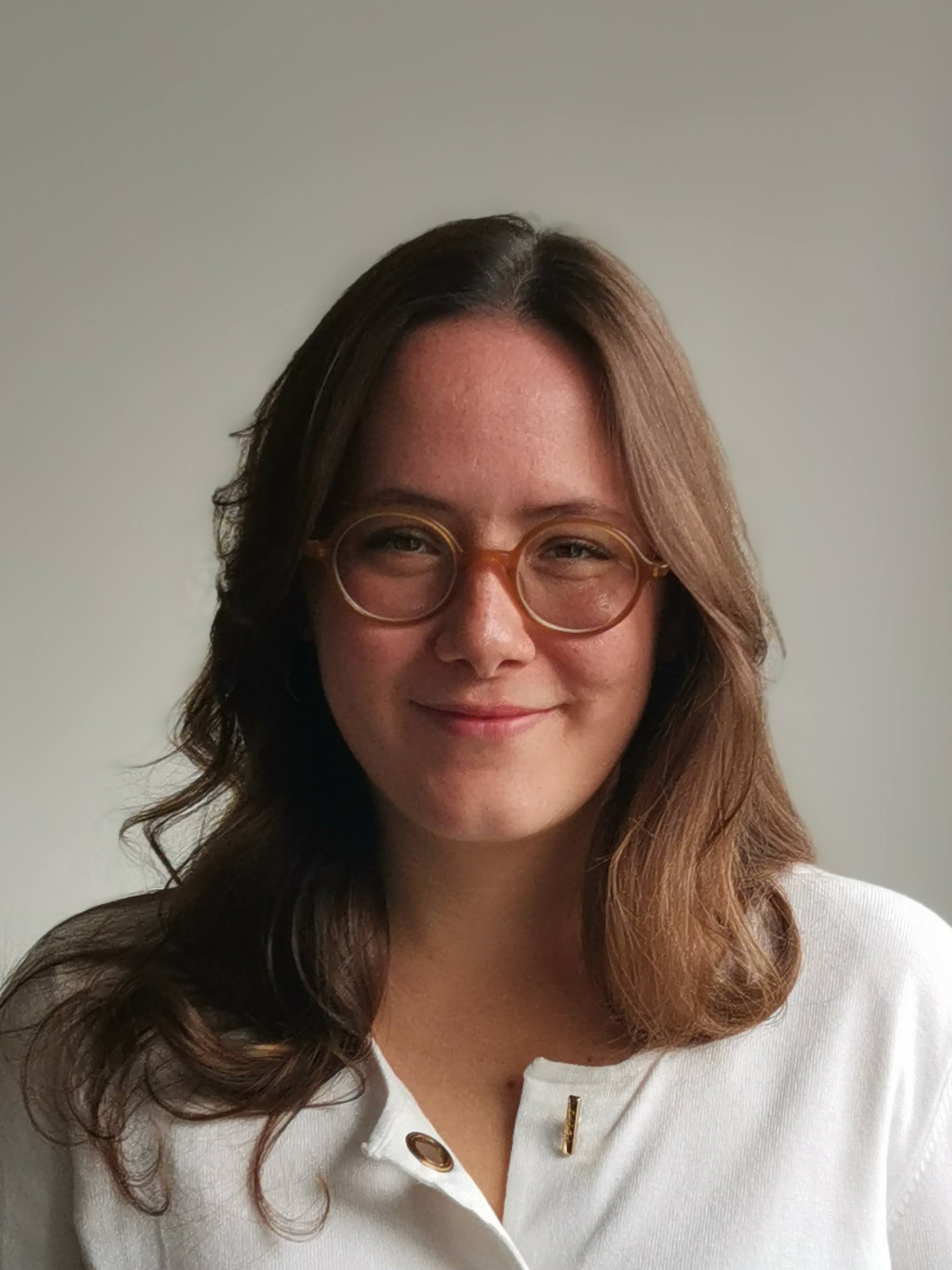}}]{Maria Flors Mor-Ruiz}
received the B.S. degree from the University of Barcelona, Spain, in 2018 and the M.S. degree from the Delft University of Technology, The Netherlands, in 2021. She is currently pursuing the Ph.D. degree with the University of Innsbruck, Austria, working on quantum communication and quantum networks. 
\end{IEEEbiography}

\vskip -2\baselineskip plus -1fil
\vspace{11pt}

\begin{IEEEbiography}[{\includegraphics[width=1in,height=1.25in,clip,keepaspectratio]{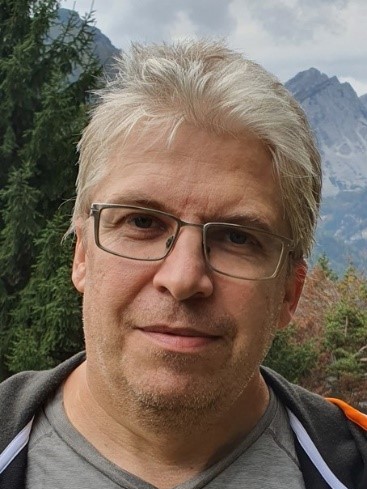}}]{Wolfgang D\"ur} is Professor for Theoretical Physics at the University of Innsbruck, leading the Quantum Communication and Quantum Networks group. He is interested in all aspects of quantum information theory, where he has contributed to various topics, including quantum communication, measurement-based quantum computation, quantum metrology, macroscopic quantum systems, and multipartite entanglement. He co-invented the W-states and the quantum repeater, which was later recognized as one of the fundamental elements for long-distance quantum communication. He worked on various aspects of quantum communication and quantum networks in the last 20 years, including protocols for entanglement purification and error correction, repeater schemes for bi- and multi-partite systems, and network architectures. 
\end{IEEEbiography}

\end{document}